\documentclass[pdflatex,sn-mathphys-num]{sn-jnl}

\usepackage{graphicx}%
\usepackage{float}%
\usepackage{multirow}%
\usepackage{amsmath,amssymb,amsfonts}%
\usepackage{amsthm}%
\usepackage{mathrsfs}%
\usepackage[title]{appendix}%
\usepackage{xcolor}%
\usepackage{textcomp}%
\usepackage{manyfoot}%
\usepackage{booktabs}%
\usepackage{algorithm}%
\usepackage{algorithmicx}%
\usepackage{algpseudocode}%
\usepackage{listings}%
\usepackage{enumitem}%
\usepackage{pgfplots}
\pgfplotsset{compat=1.18}
\usepgfplotslibrary{groupplots}

\usepackage[normalem]{ulem}
\long\def\add#1{#1}
\long\def\cut#1{}

\makeatletter
\let\table\tableorg
\let\endtable\endtableorg
\makeatother

\theoremstyle{thmstyleone}%
\theoremstyle{thmstyletwo}%

\theoremstyle{thmstylethree}%

\begin{document}

\title[SSP-KAN]{Small-scale photonic Kolmogorov-Arnold networks using standard telecom nonlinear modules}

\author*[1]{\fnm{Luca} \sur{Nogueira Cal\c{c}ado}}\email{l.cal\_ado@aston.ac.uk}

\author[1]{\fnm{Sergei K.} \sur{Turitsyn}}\email{s.k.turitsyn@aston.ac.uk}

\author[1]{\fnm{Egor} \sur{Manuylovich}}\email{e.manuylovich@aston.ac.uk}

\affil*[1]{\orgdiv{Aston Institute of Photonic Technologies (AiPT)}, 
\orgname{Aston University}, 
\orgaddress{\city{Birmingham}, \postcode{B4 7ET}, \country{United Kingdom}}}

\abstract{ Photonic neural networks promise ultrafast inference, yet most architectures rely on linear optical meshes with electronic nonlinearities, reintroducing optical-electrical-optical bottlenecks. Here we introduce small-scale photonic Kolmogorov-Arnold networks (SSP-KANs) implemented entirely with standard telecommunications components. Each network edge employs a trainable nonlinear module composed of a Mach-Zehnder interferometer, semiconductor optical amplifier, and variable optical attenuators, providing a four-parameter transfer function derived from gain saturation and interferometric mixing. Despite this constrained expressivity, SSP-KANs comprising only a few optical modules achieve strong nonlinear inference performance across classification, regression, and image recognition tasks, approaching software baselines with significantly fewer parameters. A four-module network achieves \cut{98.4\%}\add{$94.3$\% (IQR: $90.3$--$97.4$\%, 10~seeds)} accuracy on nonlinear classification benchmarks\cut{ inaccessible to linear models}\add{; a seven-module network attains $R^2 = 0.986 \pm 0.015$ on six-input regression}. Performance remains robust under realistic hardware impairments, maintaining high accuracy down to 6-bit input resolution and 14 dB signal-to-noise ratio. By using a fully differentiable physics model for end-to-end optimisation of optical parameters, this work establishes a practical pathway from simulation to experimental demonstration of photonic KANs using commodity telecom hardware.}




\maketitle

\section{Introduction}\label{sec1}
As neural networks become increasingly embedded within optical technologies - from high-capacity communications and information processing to advanced imaging and sensing (see e.g. ~\cite{Shastri2021,R01,R02,ANN01,ANN02,R08,ANN03,R04,Review001} and references therein) - there is rising demand for inference hardware capable of matching the speed and bandwidth of light itself. 
Although photonic neural networks (PNNs) promise orders-of-magnitude improvements in latency and energy efficiency by performing computation directly in the optical domain, most existing implementations follow the conventional multilayer perceptron (MLP) paradigm. In these architectures, linear optical matrix-vector multiplication,  implemented using interferometric meshes or wavelength multiplexing, is combined with electronic nonlinear activation functions. This hybrid approach reintroduces optical-electrical-optical (OEO) conversion bottlenecks, thereby limiting scalability and offsetting many of the anticipated advantages of all-optical processing~\cite{Wetzstein2020,mcmahon2023physics}. \add{Even high-performance accelerators achieving 11~TOPS throughput~\cite{Xu2021} and parallel convolutional processing on photonic tensor cores~\cite{Feldmann2021,Miscuglio2020} rely on electronic nonlinearities between optical layers.}

Kolmogorov-Arnold Networks (KANs) offer a fundamentally different computation paradigm ~\cite{Liu2024KAN}. Inspired by the Kolmogorov-Arnold representation theorem, KANs depart from the conventional node-based activation model by placing trainable univariate nonlinear functions on network edges. In contrast to multilayer perceptrons (MLPs), where fixed activation functions follow linear transformations, KANs distribute nonlinearity across connections, restructuring the computational graph itself. Recent work has demonstrated that this architecture can achieve accuracy comparable to, and in some cases exceeding, that of MLPs while requiring substantially fewer trainable parameters. Moreover, the learned edge functions admit symbolic approximation, enhancing interpretability and analytical transparency~\cite{Liu2024KAN}.

Recent results in short-reach IM/DD optical communication systems show that KAN-based equalizers can achieve superior or comparable BER performance while using substantially fewer trainable parameters than conventional FNN and CNN architectures, demonstrating reductions of up to $\sim$ 66\% \cite{chen2025kolmogorov}. These properties are particularly attractive for photonic implementation, where each nonlinear element incurs optical loss, hardware complexity, and fabrication overhead. By concentrating expressive power into a reduced number of structured nonlinear modules, KANs provide a parameter-efficient and physically compatible framework for scalable optical neural computation.

 Several recent advances provide strong motivation for photonic implementations of KANs. Fischer \emph{et al.}~\cite{Fischer2024} applied software-based KANs to nonlinear equalization in 112~Gb/s passive optical networks, demonstrating that KAN equalizers outperform convolutional neural networks at comparable computational complexity. These results show that KAN architectures are effective for high-speed optical signal processing and motivate their direct realization in the optical domain. Peng \emph{et al.}~\cite{Peng2024} reported the first photonic KAN implementation using ring-assisted Mach-Zehnder interferometers (RAMZI), using free-carrier dispersion in silicon microring resonators to create tunable nonlinear transfer functions. Their integrated silicon photonic architecture achieved 98\% accuracy on MNIST handwritten digit recognition while reducing the energy-area product by approximately 65-fold compared to conventional MZI-based optical neural networks.

More recently, Stroev and Berloff~\cite{Stroev2025} introduced a theoretical framework for all-optical KANs implemented on spatial light modulator platforms, showing that trainable univariate nonlinear functions can arise from structural interference effects without relying on conventional nonlinear optical materials.

Despite these advances, two fundamental challenges remain. First, it is unclear whether the constrained nonlinear transfer functions available from practical optical components (often defined by only a small number of tunable physical parameters) can achieve expressive capability comparable to software-based activation functions such as B-splines, which may involve hundreds of learnable coefficients per edge. Second, and more critically for real-world deployment, it remains an open question whether photonic KANs can be implemented using off-the-shelf telecommunications components that are mass-produced, well-characterized, and optimized for the spectral bands and power levels of fibre-optic systems.

\cut{Previous photonic KAN demonstrations have relied on specialized resonator geometries or free-space spatial light modulator platforms, and conventional photonic neural networks face different but related constraints: photonic}
\add{Conventional photonic neural networks face related but distinct constraints: photonic} multilayer perceptrons perform linear matrix-vector multiplication optically but require electronic nonlinear activations between layers, while photonic reservoir computers exploit intrinsic nonlinear dynamics but use fixed, untrained nonlinearities with only a linear readout optimized.
\add{Previous photonic KAN demonstrations have relied on specialized resonator geometries or engineered nonlinear waveguides.}
\cut{SSP-KAN occupies a different position.}
\cut{Each optical nonlinearity is independently trainable through physically grounded parameters, the nonlinear transformation occurs on every network edge rather than at nodes, and the entire system is optimized end-to-end through a differentiable physics model.}
\add{SSP-KAN differs from prior photonic KAN architectures in hardware accessibility. The RAMZI approach~\cite{Peng2024} requires custom silicon photonic fabrication with precise resonance tuning; the SPM-based design~\cite{sozos_photonic_2026} requires engineered nonlinear waveguides tailored at the foundry level. SSP-KAN uses only commercially available, fibre-pigtailed telecom components and needs no custom fabrication. The nonlinearity is SOA gain saturation, a broadband travelling-wave effect that operates across the full C-band. Embedding the SOA within an MZI adds an interferometric phase degree of freedom; the resulting four-parameter module produces a family of nonlinear transfer functions from saturable gain combined with interferometric mixing.}
This per-edge trainability is what enables compositional depth to compensate for the limited expressivity of each individual module, while the use of commodity telecom hardware lowers the barriers to experimental realization.

Here we demonstrate that physically constrained photonic nonlinearities can act as trainable activation functions within Kolmogorov–Arnold Networks, enabling nonlinear inference using standard telecommunications components. Specifically, we show that Mach-Zehnder interferometers combined with semiconductor optical amplifiers and variable optical attenuators (MZI-VOA-SOA-VOA modules) can function as compact, trainable nonlinear units for photonic KANs, with each module controlled by only four physical parameters. Our approach differs from prior implementations based on specialized resonator geometries or custom optical platforms, in two key aspects: (i) it uses SOA gain saturation as a primary (native to telecom) nonlinearity that does not require bespoke fabrication; (ii) the nonlinear phase shift from SOA amplification is employed as a source of nonlinearity  by placing the SOA inside an MZI. We have developed a fully differentiable physics-based model that enables end-to-end gradient optimization of all optical parameters, including injection current, attenuation coefficients, and interferometric phase.

\cut{Despite the limited per-module parameter count, SSP-KAN comprising only 4-8 optical modules achieve strong nonlinear inference performance. Using commercially specified SOA parameters (Thorlabs BOA1554P), we obtain 99.1\% accuracy on nonlinear classification benchmarks and a coefficient of determination $R^2 = 0.977$ on multivariate regression.} \add{Using commercially specified SOA parameters (Thorlabs BOA1554P), a four-module [2,2] SSP-KAN achieves $94.3$\% test accuracy (IQR: $90.3$--$97.4$\%, 10~seeds) on Two Moons nonlinear classification, and a seven-module [6,1,1] network attains $R^2 = 0.986 \pm 0.015$ on six-input yacht hydrodynamics regression.} Scaling studies on MNIST image classification further demonstrate competitive performance, reaching \cut{92.7}\add{$93.9 \pm 0.2$}\% accuracy\add{ (10~seeds)}. Systematic robustness analysis under optical noise and finite-resolution input quantization confirms stable operation across realistic impairment regimes.

Overall, we introduce a SSP-KAN design framework grounded entirely in standard telecommunications components. The approach is inherently modular and can be extended to alternative nonlinear devices and photonic architectures, providing a foundation for telecom-grade implementations of trainable optical neural networks. Rather than pursuing asymptotic universality, this work demonstrates that physically constrained, low-parameter photonic KANs are sufficient for meaningful nonlinear inference in realistic settings. Such architectures are particularly well suited to structured computational tasks, including physics-informed modeling, low-dimensional regression, symbolic discovery, latency-critical edge inference, and applications where interpretability and hardware efficiency are paramount.

\section{Results}\label{sec2}
\subsection{SSP-KAN architecture}\label{sec:results:architecture}

\begin{figure}[H]
\centering
\includegraphics[width=\textwidth]{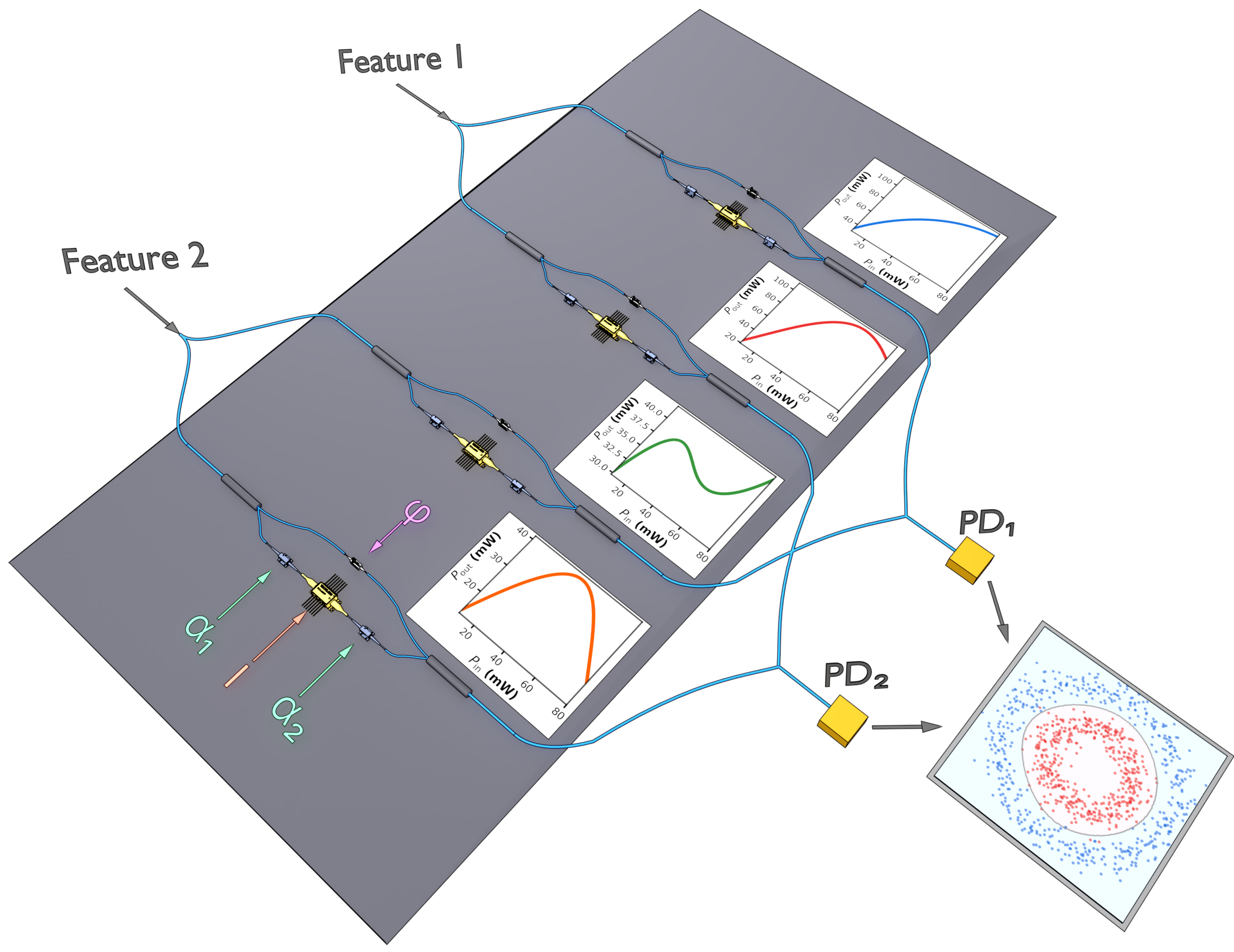}
\caption{SSP-KAN [2,2] network. Two input features are encoded as optical power and split to four MZI-VOA-SOA-VOA modules on a fibre-coupled breadboard. Each module (detail, lower left) has four trainable parameters: input attenuation~$\alpha_1$, injection current~$I$, output attenuation~$\alpha_2$, and interferometric phase~$\phi$. Insets show the learned transfer function $P_{\text{out}}(P_{\text{in}})$ of each module. The per-edge outputs are summed on two photodetectors (PD$_1$, PD$_2$), producing the nonlinear decision boundary on the concentric rings task (right).}
\label{fig:architecture}
\end{figure}

The main idea of SSP-KAN is to replace software-defined activation functions with physically realized optical transfer functions, placing a dedicated MZI-VOA-SOA-VOA module on each network edge.
Figure~\ref{fig:architecture} shows a minimal [2,2] network with two input nodes and two output nodes, requiring four MZI-VOA-SOA-VOA modules. Each output is computed as a sum of the nonlinearly-transformed inputs:
\begin{equation}
\cut{y_j = \sum_{i=1}^{n_{\text{in}}} \phi_{i,j}(x_i)}
\add{y_j = \sum_{i=1}^{n_{\text{in}}} \varphi_{i,j}(x_i)}
\label{eq:kan_layer}
\end{equation}
where \cut{$\phi_{i,j}$} \add{$\varphi_{i,j}$} represents the transfer function of the module on edge $(i,j)$. This is the key structural difference from multilayer perceptrons: rather than applying a fixed nonlinearity after a weighted sum, each edge independently transforms its input through a physically distinct optical transfer function. The summation itself can be performed through incoherent power combining on a photodetector, requiring no electronic computation within the network.

Each module (detail in Fig.~\ref{fig:architecture}) consists of a Mach-Zehnder interferometer with a semiconductor optical amplifier and variable optical attenuators in one arm, providing a tunable nonlinear transfer function controlled by four trainable parameters. The SOA injection current $I$ sets both the gain magnitude and the degree of saturation nonlinearity, ranging from approximately linear response at low currents to pronounced gain compression at high currents. The input attenuation $\alpha_1$ controls the optical power entering the SOA, thereby setting the operating point on the gain saturation curve: low attenuation drives the SOA into saturation with strong nonlinearity, while high attenuation keeps it in the linear amplification regime. The output attenuation $\alpha_2$ scales the amplified arm contribution to the interference without affecting the saturation dynamics. The interferometer phase $\phi$ selects which portion of the MZI cosine-squared fringe pattern reaches the output, choosing between constructive and destructive interference regimes. Together, these four parameters per module define a family of transfer functions that spans a rich space of nonlinear input-output relationships (insets in Fig.~\ref{fig:architecture}). All simulations use manufacturer-specified parameters for a commercially available SOA (Thorlabs BOA1554P; see Methods for the full physical model).

\subsection{Nonlinear classification on Two Moons}\label{sec:results:moons}

To establish that SSP-KAN can learn nonlinear decision boundaries, we evaluate on the Two Moons dataset, a standard binary classification benchmark where two interleaving half-circles cannot be separated by any linear classifier. The dataset comprises 1,000 training and 1,000 test samples with Gaussian noise ($\sigma = 0.1$), scaled to the optical power range 10--80~mW. We train a single-layer [2,2] network comprising 4~MZI-VOA-SOA-VOA modules (16~trainable parameters) and compare against a linear baseline and a software KAN baseline implemented using pyKAN~\cite{Liu2024KAN} with grid size $G{=}1$ and B-spline activation functions.

\begin{figure}[H]
\centering
\includegraphics[width=\textwidth]{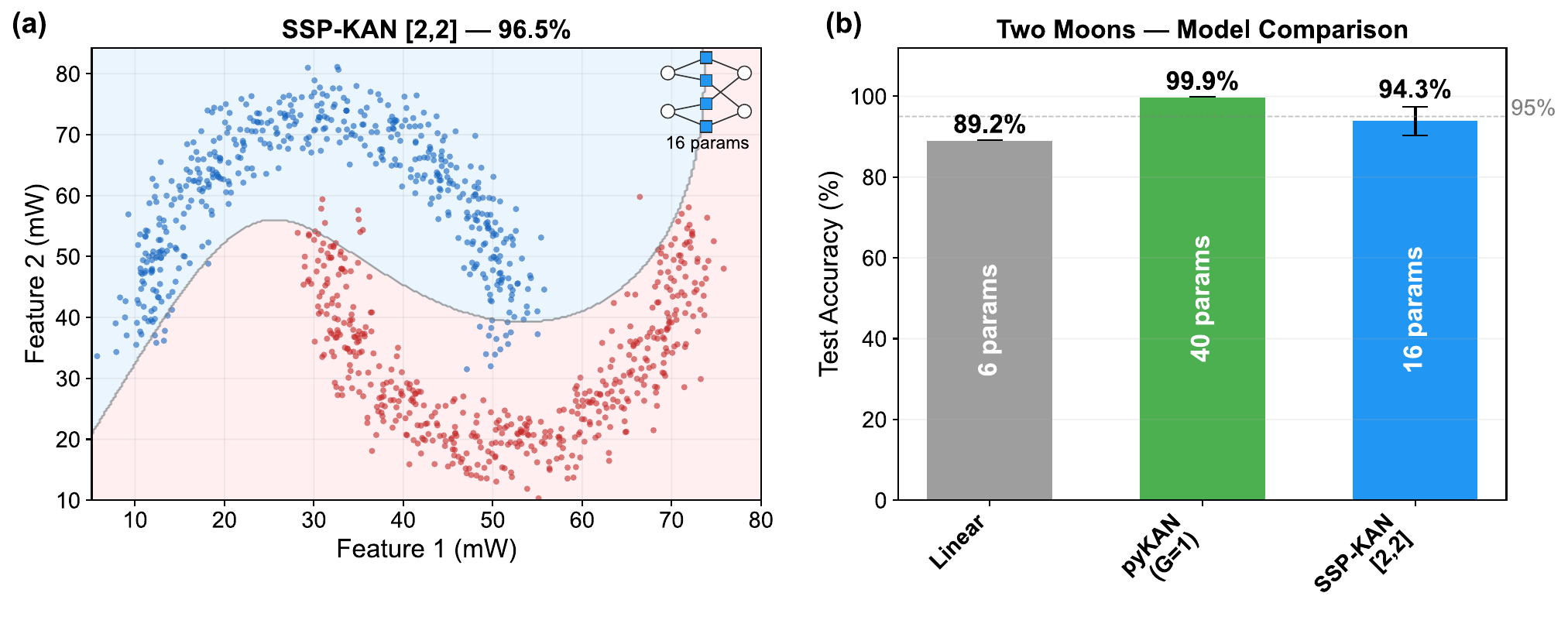}
\caption{Two Moons classification with SSP-KAN~[2,2]. \textbf{(a)}~Learned decision boundary (\cut{98.4\%}\add{$94.3$\% (IQR: $90.3$--$97.4$\%), 10~seeds;} test accuracy, 16~parameters). The network separates the two crescents using a smooth nonlinear boundary generated by four MZI-VOA-SOA-VOA modules. \textbf{(b)}~Test accuracy and parameter count across models. The linear baseline (89.2\%, 6~parameters) confirms the task is intrinsically nonlinear. pyKAN ($G{=}1$, 40~parameters) reaches 99.9\%; SSP-KAN achieves \cut{comparable accuracy (98.4\%) with fewer than half the parameters and using only physically realizable optical activation functions}\add{$94.3$\% with 16~parameters using physically realisable optical activation functions}. Inset: [2,2] network schematic showing 2~inputs, 4~modules, and 2~outputs.}
\label{fig:moons_clean}
\end{figure}

Figure~\ref{fig:moons_clean}a shows the learned decision boundary for the [2,2] architecture. The network achieves \cut{98.4\%}\add{$94.3$\% (IQR: $90.3$--$97.4$\%, 10~seeds)} test accuracy, separating the two crescents with a smooth nonlinear boundary that follows the data geometry. The linear baseline reaches only 89.2\% (Fig.~\ref{fig:moons_clean}b), confirming the intrinsically nonlinear nature of the task. The software KAN baseline (pyKAN, $G{=}1$) achieves 99.9\% with 40~parameters; SSP-KAN closes to within \cut{1.5}\add{5.6}~percentage points using 16~parameters and a transfer function family constrained to the physics of MZI-VOA-SOA-VOA modules. This \cut{small} accuracy gap \cut{demonstrates}\add{reflects} that the nonlinear response of saturated semiconductor optical amplifiers, shaped by the four trainable parameters per edge, provides sufficient functional diversity to approximate the B-spline activation functions used in software KANs\add{, though the periodic MZI transfer function creates multiple local minima that widen the gap relative to single-seed results (best seed: 98.9\%, worst: 88.1\%)}. A deeper [2,2,2] architecture with 8~modules \cut{marginally improves no-noise accuracy}\add{substantially improves accuracy to $99.1$\% (IQR: $99.0$--$99.1$\%, 10~seeds)} (Supplementary Section~A)\cut{; however, multi-layer}\add{, demonstrating that compositional depth compensates for limited per-module expressivity. Multi-layer} operation introduces coherent phase accumulation at intermediate summation nodes, which is analyzed in Supplementary Section~B.

\begin{figure}[H]
\centering
\includegraphics[width=\textwidth]{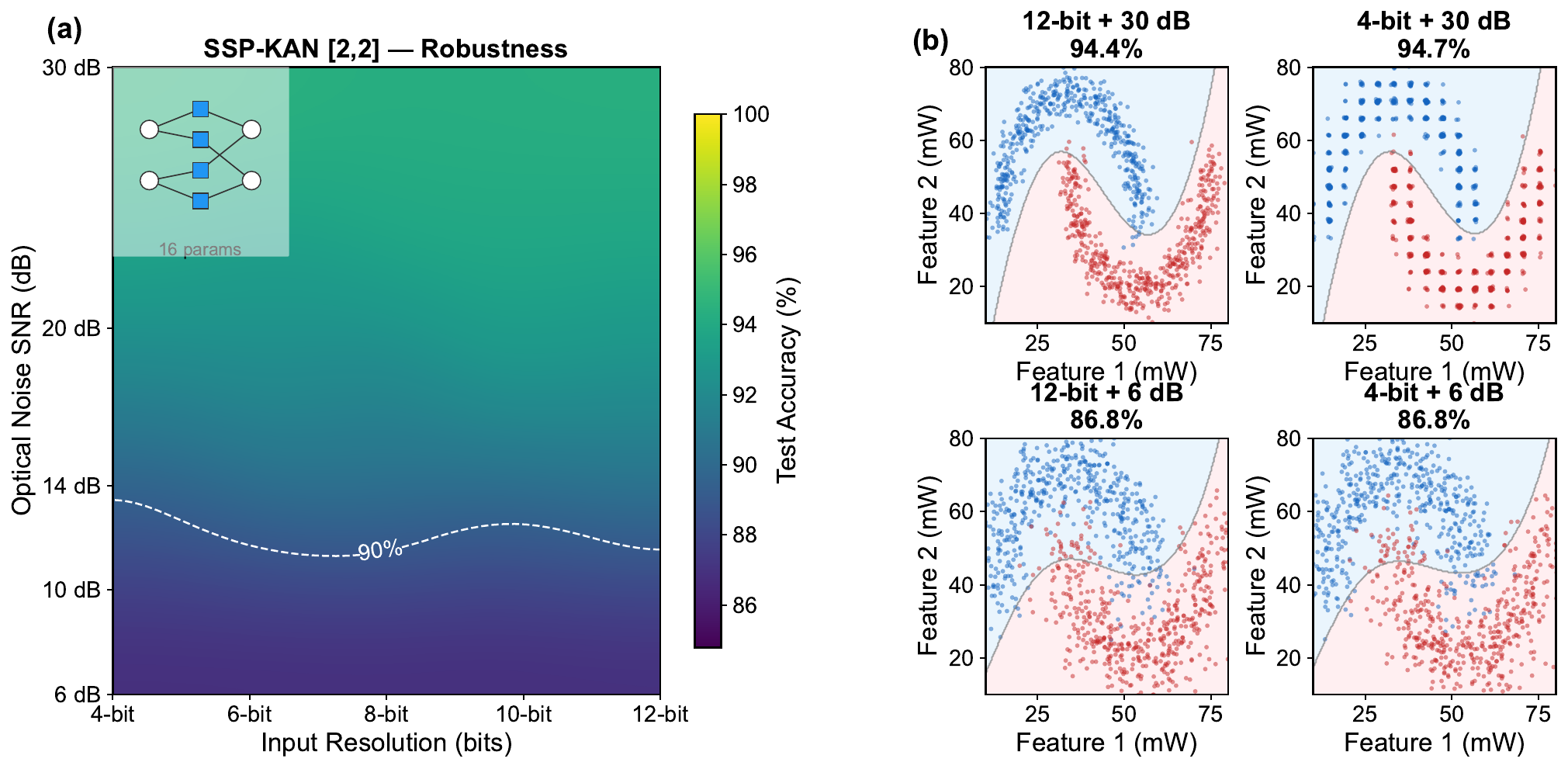}
\caption{Hardware robustness of SSP-KAN~[2,2] on Two Moons.
\textbf{(a)}~Test accuracy as a function of input DAC resolution (12-bit to 4-bit) and optical noise (SNR 30~dB to 6~dB). The smooth colour gradient confirms that quantization has negligible impact, while performance degrades vertically with decreasing SNR. The 90\% and 95\% contour lines indicate the practical operating boundary. Inset: [2,2] network schematic (16~parameters). \textbf{(b)}~Decision boundaries at four corner conditions. Top row: 30~dB noise at 12-bit (\cut{95.4}\add{94.4}\%) and 4-bit (\cut{94.5}\add{94.7}\%) resolution. Bottom row: 6~dB noise at 12-bit (\cut{87.9}\add{86.8}\%) and 4-bit (\cut{87.8}\add{86.8}\%). The boundary shape is preserved across
quantization levels but simplifies under severe noise.}
\label{fig:moons_robustness}
\end{figure}

Having established classification performance under ideal conditions, we next evaluate robustness to the two dominant hardware impairments in photonic systems: finite digital-to-analogue converter (DAC) resolution for input encoding and optical noise from amplified spontaneous emission (ASE). Figure~\ref{fig:moons_robustness}a shows classification accuracy for the [2,2] network across input quantization levels (12-bit down to 4-bit) and optical SNR conditions (30~dB down to 6~dB). Quantization has negligible effect on performance: at any given SNR, accuracy is nearly constant across all tested resolutions. This implies that inexpensive, low-resolution DACs are sufficient for input encoding.

Optical noise, by contrast, produces measurable degradation. At SNR~$=$~30~dB accuracy is \cut{above 95\%, dropping to approximately 97\% at 20~dB, 96\% at 14~dB, and 94\% at 10~dB}\add{approximately 94\%, dropping to 94\% at 20~dB, 92\% at 14~dB, and 90\% at 10~dB}. Under extreme noise (SNR~$=$~6~dB), the network retains approximately 87\% accuracy, falling below the linear baseline (89.2\%). At sufficiently high noise levels the nonlinear decision boundary becomes a liability: the optimiser converges to a coarser separation that is less effective than a simple linear cut. The decision boundaries under extreme conditions (Fig.~\ref{fig:moons_robustness}b) illustrate this directly. Under 6~dB noise, the boundary approximates a broad diagonal rather than tracking the crescent geometry. This crossover defines a practical operating threshold for the [2,2] architecture: for SNR~$\geq$~10~dB, the nonlinear model consistently outperforms the linear baseline, while below 10~dB, simpler models may be preferable. Fig.~\ref{fig:moons_robustness}b confirms the quantization finding: the 4-bit boundary (\cut{87.8}\add{86.8}\%) is visually and numerically indistinguishable from the 12-bit boundary (\cut{87.9}\add{86.8}\%) at the same noise level. For realistic deployment conditions (8-bit DAC, SNR~$\geq$~14~dB), the [2,2] architecture maintains accuracy above \cut{96}\add{91}\%, comfortably exceeding the linear baseline.

\subsection{Regression performance on yacht hydrodynamics}\label{sec:results:regression}

To evaluate SSP-KAN on a higher-dimensional task, we use the UCI Yacht Hydrodynamics dataset~\cite{UCI}, derived from Delft ship hydrodynamics experiments~\cite{Gerritsma1981}. The dataset contains 308 samples, each described by six hull geometry features (longitudinal position of the centre of buoyancy, prismatic coefficient, length-displacement ratio, beam-draught ratio, length-beam ratio, and Froude number) with the prediction target being residuary resistance per unit weight of displacement. This regression task tests whether SSP-KAN can learn complex multivariate functions from six-dimensional inputs using a small number of optical modules.

\begin{figure}[H]
\centering
\includegraphics[width=.7\textwidth]{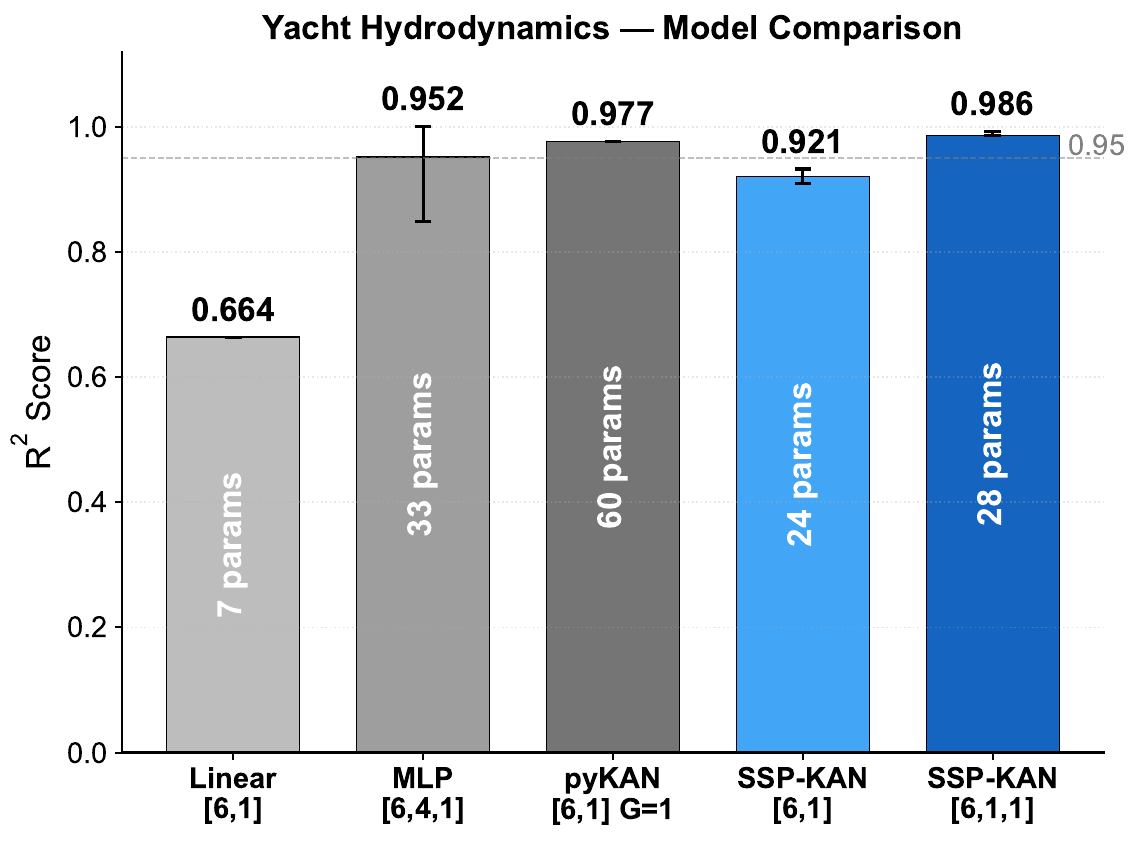}
\caption{Regression performance on yacht hydrodynamics. Test $R^2$ scores \add{(mean $\pm$ std, 10~seeds)} and parameter counts for linear regression, MLP, software KAN (pyKAN, $G{=}1$), and SSP-KAN architectures [6,1] and [6,1,1]. SSP-KAN~[6,1,1] \cut{matches}\add{exceeds} pyKAN ($R^2 = 0.977$) with 28~parameters versus 60, demonstrating that compositional depth compensates for the constrained per-edge activation family.}
\label{fig:yacht_comparison}
\end{figure}

We evaluate two architectures: a single-layer [6,1] network with 6~MZI-VOA-SOA-VOA modules (24~trainable parameters) and a two-layer [6,1,1] network with 7~modules (28~parameters). Figure~\ref{fig:yacht_comparison} compares both against baseline models. A linear regression baseline achieves $R^2 = 0.664$, confirming the nonlinear nature of the hull-resistance relationship. A multilayer perceptron (MLP) with architecture [6$\rightarrow$4$\rightarrow$1] and 33~parameters achieves $R^2 = \cut{0.989}\add{0.952 \pm 0.103}$\add{ (10~seeds)}, while a software KAN (pyKAN, [6,1] with grid size $G{=}1$, 60~parameters) reaches $R^2 = 0.977$. The single-layer SSP-KAN~[6,1] achieves $R^2 = \cut{0.869}\add{0.921 \pm 0.017}$\add{ (10~seeds)}: a substantial improvement over the linear baseline, but limited by the fact that each of its six modules computes a univariate function of a single input feature, with no capacity for feature interaction. Adding a second layer improves performance dramatically: SSP-KAN~[6,1,1] reaches $R^2 = \cut{0.977, matching}\add{0.986 \pm 0.015}$, \cut{matching}\add{exceeding} pyKAN \cut{exactly }while using fewer than half the parameters (28~vs~60). The second-layer module composes the six first-layer outputs into a single nonlinear function, enabling the network to capture multivariate interactions that no sum of univariate functions can represent.

\begin{figure}[H]
\centering
\includegraphics[width=\textwidth]{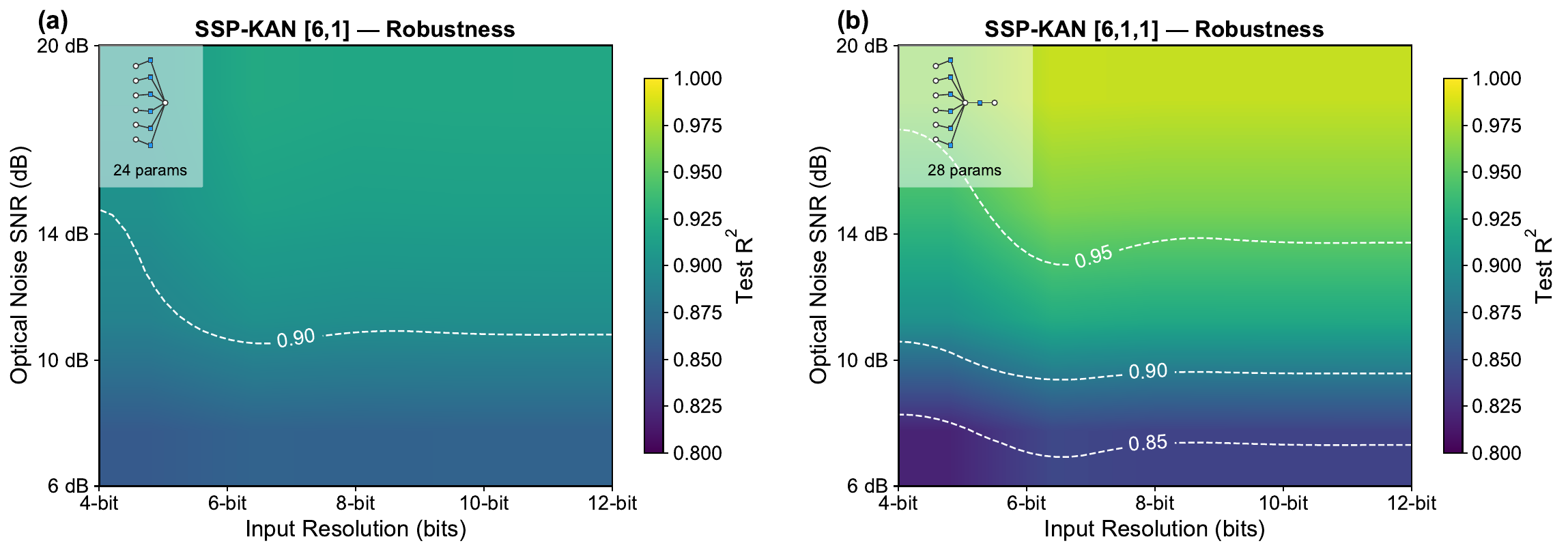}
\caption{Hardware robustness of SSP-KAN on yacht hydrodynamics regression. Test $R^2$ as a function of input DAC resolution (12-bit to 4-bit) and optical noise (SNR 30~dB to 6~dB) for \textbf{(a)}~the single-layer [6,1] architecture (6~modules, 24~parameters) and \textbf{(b)}~the two-layer [6,1,1] architecture (7~modules, 28~parameters). Contour lines mark $R^2$ thresholds. The [6,1] panel shows a compressed performance range ($R^2 \approx \cut{0.78}\add{0.86}$--$\cut{0.87}\add{0.92}$), while [6,1,1] spans a wider range ($R^2 \approx \cut{0.78}\add{0.82}$--$\cut{0.98}\add{0.99}$), reflecting both higher peak performance and greater noise sensitivity.} 
\label{fig:yacht_robustness}
\end{figure}

Robustness to hardware impairments is assessed across SNR levels from 6~dB to 30~dB \cut{conditions }and DAC resolutions from 4-bit to 12-bit (Fig.~\ref{fig:yacht_robustness}). Figure~\ref{fig:yacht_robustness}a,b shows $R^2$ as a function of input resolution at each noise level for the [6,1] and [6,1,1] architectures, respectively. As with Two Moons classification, input quantization has minimal effect: for the [6,1,1] architecture at SNR~=~30~dB, $R^2$ remains between \cut{0.975 and 0.977}\add{0.98 and 0.99} from 12-bit down to 8-bit, with a \cut{modest }decrease to \cut{approximately 0.95}\add{$R^2 \approx 0.96$} at 4-bit. Optical noise produces more significant degradation, reducing [6,1,1] performance from $R^2 \approx \cut{0.977}\add{0.98}$ at SNR~=~30~dB \cut{conditions }to approximately \cut{0.95}\add{0.98} at SNR~$=$~20~dB, \cut{0.91}\add{0.96} at 14~dB, \add{0.92 at 10~dB, }and \cut{0.78}\add{0.84} at 6~dB.

\cut{A notable feature of} Fig.~\ref{fig:yacht_robustness} \add{shows a clear} contrast between the two panels. The [6,1] architecture (Fig.~\ref{fig:yacht_robustness}a) shows performance compressed into a narrow range between $R^2 \approx \cut{0.78}\add{0.86}$ and \cut{0.87}\add{0.92} across the entire parameter space. The [6,1,1] architecture (Fig.~\ref{fig:yacht_robustness}b) spans a much wider range, from $R^2 \approx \cut{0.98}\add{0.99}$ at 30~dB down to \cut{0.78}\add{0.82} at 6~dB, with contour lines indicating where each performance threshold is crossed. This contrast directly visualizes the depth-robustness trade-off: the deeper network achieves substantially higher performance under mild impairment but is more sensitive to perturbation, as the second-layer module amplifies noise that has already propagated through the first layer. The shallower [6,1] network, by contrast, has less capacity to lose. \cut{Both architectures converge to similar performance ($R^2 \approx 0.78$) under extreme noise, suggesting a noise floor set by the intrinsic difficulty of the regression task rather than by architectural capacity.} \add{Under extreme noise (6~dB), the [6,1] network ($R^2 \approx 0.86$) outperforms [6,1,1] ($R^2 \approx 0.82$), indicating that the deeper architecture's greater noise sensitivity outweighs its capacity advantage at this SNR.} Under combined impairments representative of realistic deployment (8-bit DAC, SNR~$=$~14~dB), the [6,1,1] architecture maintains $R^2 \approx \cut{0.91}\add{0.96}$, well above the linear baseline ($R^2 = 0.664$). A mild increase in quantization sensitivity is visible at 4-bit for the [6,1,1] network under lower SNR conditions, suggesting that 6-bit DAC resolution represents a practical lower bound for the deeper architecture.
\subsection{Image classification on MNIST and Fashion-MNIST}\label{sec:results:mnist}

To evaluate SSP-KAN on high-dimensional inputs representative of practical inference tasks, we tested on MNIST handwritten digit classification (784-dimensional input, 10 classes) and Fashion-MNIST clothing classification (same dimensionality, 10 classes). MNIST is a direct comparison point with the D-RAMZI photonic KAN of Peng \emph{et al.}~\cite{Peng2024}, while also testing whether the MZI-VOA-SOA-VOA transfer function can handle image-scale inputs. Fashion-MNIST is a substantially harder benchmark: its classes share more visual similarity (e.g.\ shirts, coats, and pullovers occupy overlapping regions of pixel space), making it a more demanding test of whether optical nonlinearity provides meaningful benefit over linear classification.

We evaluated three SSP-KAN architectures of increasing capacity: a single-layer [784,10] network with 7,840 MZI-VOA-SOA-VOA modules, a two-layer [784,10,10] network with 7,940 modules, and a wider two-layer [784,20,10] network with 15,880 modules. All models were trained with identical hyperparameters (AdamW optimiser, weight decay $10^{-3}$, 150 epochs; see Methods). A linear classifier trained under the same conditions is the baseline, isolating the contribution of optical nonlinearity from the effect of increased model capacity.

\begin{figure}[H]
\centering
\includegraphics[width=\textwidth]{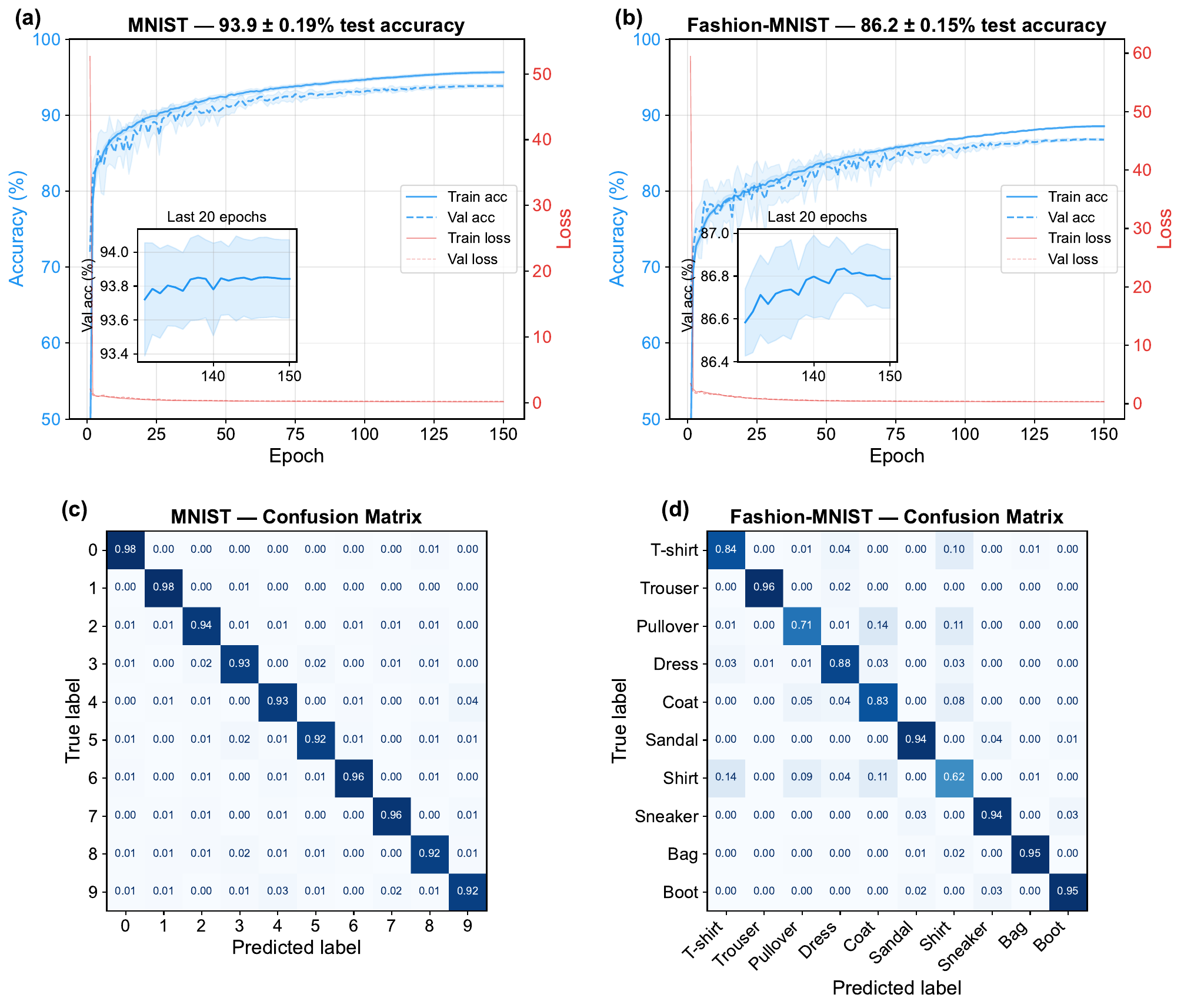}
\caption{Image classification results. Training curves showing accuracy and loss for (a)~MNIST and (b)~Fashion-MNIST using the [784,20,10] architecture. Insets show the final training epochs at expanded scale. Confusion matrices for the [784,20,10] model on (c)~MNIST and (d)~Fashion-MNIST, showing that classification errors concentrate among visually similar classes.}
\label{fig:mnist}
\end{figure}

On MNIST, the linear baseline achieves 91.0\% accuracy, consistent with the well-known near-linear separability of handwritten digits in pixel space. \cut{SSP-KAN [784,10] reaches 91.8\%, SSP-KAN [784,10,10] reaches 92.0\%, and SSP-KAN [784,20,10] reaches 92.7\%, a gain of 1.7 percentage points over the linear model}\add{SSP-KAN [784,20,10] reaches $93.9 \pm 0.2$\% (10~seeds), a gain of 2.9 percentage points over the linear model} (Fig.~\ref{fig:mnist}). On Fashion-MNIST, where the classification task is genuinely nonlinear, \cut{the advantage of optical nonlinearity becomes more pronounced. The}\add{the} linear baseline drops to 83.3\%, while SSP-KAN [784,20,10] achieves \cut{85.7}\add{$86.2 \pm 0.2$}\%\cut{, widening the gap to}\add{, maintaining the same} \cut{2.4}\add{2.9} percentage point\add{s} \add{gain over the linear model} (Fig.~\ref{fig:mnist}b). \cut{This pattern is consistent across all architectures: the SSP-KAN improvement over the linear model is systematically larger on Fashion-MNIST than on MNIST, confirming that the}\add{The consistent margin across both tasks confirms that} MZI-VOA-SOA-VOA transfer functions provide \cut{greater} benefit on tasks where nonlinear feature interactions are essential for classification.

The confusion matrices (Fig.~\ref{fig:mnist}c,d) reveal that SSP-KAN errors are concentrated among visually similar classes. On MNIST, the most common confusions occur between digits that share structural features (e.g.\ 4/9, 3/5, 7/9). On Fashion-MNIST, the dominant error mode is confusion among upper-body garments (shirt, coat, pullover, and T-shirt/top), which differ primarily in subtle textural features rather than gross shape. These error patterns are consistent with the limited expressivity of four-parameter optical transfer functions: the network captures coarse nonlinear features effectively but cannot resolve fine-grained distinctions that would require higher-order activation function complexity.

\section{Methods}\label{sec11}

\subsection{Kolmogorov-Arnold Networks}

The Kolmogorov-Arnold representation theorem states that any multivariate continuous function $f: [0,1]^n \to \mathbb{R}$ can be exactly represented as a superposition of continuous univariate functions and addition:
\begin{equation}
\cut{f(x_1, \ldots, x_n) = \sum_{q=0}^{2n} \Phi_q \left( \sum_{p=1}^{n} \phi_{q,p}(x_p) \right)}
\add{f(x_1, \ldots, x_n) = \sum_{q=0}^{2n} \Phi_q \left( \sum_{p=1}^{n} \varphi_{q,p}(x_p) \right)}
\label{eq:kolmogorov}
\end{equation}
where \cut{$\phi_{q,p}: [0,1] \to \mathbb{R}$} \add{$\varphi_{q,p}: [0,1] \to \mathbb{R}$} and $\Phi_q: \mathbb{R} \to \mathbb{R}$ are continuous univariate functions~\cite{Liu2024KAN}. This theorem implies that high-dimensional function approximation can be reduced to learning one-dimensional functions, fundamentally differing from the multilayer perceptron (MLP) approach where fixed nonlinearities are applied at nodes.

KANs use this representation by placing learnable activation functions on network edges rather than fixed activations at nodes. In a standard MLP, the transformation between layers takes the form $\mathbf{y} = \sigma(\mathbf{W}\mathbf{x} + \mathbf{b})$, where $\sigma$ is a fixed nonlinearity (e.g., ReLU or sigmoid) applied element-wise after a linear transformation. In contrast, a KAN layer computes:
\begin{equation}
\cut{y_j = \sum_{i=1}^{n_{\text{in}}} \phi_{i,j}(x_i)}
\add{y_j = \sum_{i=1}^{n_{\text{in}}} \varphi_{i,j}(x_i)}
\label{eq:kan_layer_methods}
\end{equation}
This architectural difference has significant implications for photonic implementation. In conventional photonic neural networks, the linear matrix-vector multiplication $\mathbf{W}\mathbf{x}$ is naturally implemented using interference in Mach-Zehnder interferometer meshes or wavelength-division multiplexing, but the nonlinear activation $\sigma(\cdot)$ typically requires optical-to-electrical conversion, electronic processing, and electrical-to-optical conversion~\cite{Wetzstein2020}. This OEO bottleneck limits throughput and increases power consumption.

The KAN architecture inverts this challenge: rather than requiring a single powerful nonlinearity after linear mixing, KANs require many weaker nonlinearities applied independently to each signal before linear summation. Optical summation (combining beams on a photodetector or through interference) is straightforward, while tunable optical nonlinearities, though constrained in functional form, are achievable through various physical mechanisms including saturable absorption, gain saturation, and nonlinear phase shifts~\cite{Sobhanan2022}.

In software implementations, the univariate functions \cut{$\phi_{i,j}$} \add{$\varphi_{i,j}$} are typically parameterized as B-splines with learnable control points, providing smooth, flexible function approximation with $G + k$ parameters per edge, where $G$ is the grid size and $k$ is the spline order~\cite{Liu2024KAN}. A key question for photonic KANs is whether the more constrained transfer functions available from optical components, with only a handful of tunable parameters, can provide sufficient expressivity for practical tasks.

\subsection{MZI-VOA-SOA-VOA optical module}

Our photonic KAN implementation uses a composite optical module consisting of a Mach-Zehnder interferometer (MZI) with a semiconductor optical amplifier (SOA) and variable optical attenuators (VOAs) in one arm. This configuration provides a tunable nonlinear transfer function with four trainable parameters: the SOA injection current $I$, two VOA attenuation coefficients $\alpha_1$ and $\alpha_2$, and the interferometer phase $\phi$.


\subsection{Module Architecture}

The MZI splits the input optical power $P_0$ equally between two arms via a 50/50 coupler (Fig.~\ref{fig:architecture}). The lower arm contains VOA$_1$, followed by the SOA, followed by VOA$_2$. The upper arm provides a reference path with adjustable phase $\phi$. The signals recombine at a second 50/50 coupler, producing interference that shapes the output.

The power entering the SOA is:
\begin{equation}
P_{\text{SOA,in}} = \alpha_1 \cdot \frac{P_0}{2}
\label{eq:power_division}
\end{equation}
where $\alpha_1 = 10^{-A_1/10}$ is the linear transmission corresponding to VOA$_1$ attenuation $A_1$ in dB. 

Each component serves a distinct purpose: the injection current $I$ controls the SOA small-signal gain $h_0$; VOA$_1$ sets the operating point on the gain saturation curve by controlling the power entering the SOA; VOA$_2$ scales the amplitude of the SOA arm output independently of the saturation dynamics; and the phase $\phi$ determines the interference condition between the two arms.

\subsection{SOA gain dynamics}

We model the SOA using the Agrawal--Olsson formalism, where the integrated gain $h = \int_0^L g(z)\,dz$ evolves according to the rate equation: \begin{equation} \frac{dh}{dt} = \frac{h_0 - h}{\tau_c} - \frac{P_{\text{in}}}{E_{\text{sat}}}(e^h - 1) \label{eq:rate_equation} \end{equation} where $h_0$ is the small-signal integrated gain, $\tau_c$ is the carrier lifetime, and $E_{\text{sat}}$ is the saturation energy. At steady state ($dh/dt = 0$), this becomes: \begin{equation} h = h_0 - \frac{P_{\text{in}}}{P_{\text{sat}}}(e^h - 1) \label{eq:steady_state} \end{equation} where $P_{\text{sat}} = E_{\text{sat}}/\tau_c$ is the saturation power. This transcendental equation is solved numerically by Newton--Raphson iteration: \begin{equation} h_{n+1} = h_n - \frac{h_n - h_0 + (P_{\text{in}}/P_{\text{sat}})(e^{h_n} - 1)}{1 + (P_{\text{in}}/P_{\text{sat}})\,e^{h_n}} \label{eq:newton} \end{equation} initialized at $h_0$, with three iterations sufficient for convergence across the operating range. The power gain is $G = e^h$.

The small-signal gain $h_0$ depends on the injection current $I$. Above the transparency current $I_{\text{tr}}$, we use a calibrated relationship: \begin{equation} h_0(I) = \kappa \left(\frac{I}{I_{\text{tr}}} - 1\right) \label{eq:h0_current} \end{equation} where $h_{0,\text{max}} = \ln G_{\text{max}}$ and the gain coefficient $\kappa = h_{0,\text{max}}/(I_{\text{max}}/I_{\text{tr}} - 1)$ is determined from device specifications. For our simulations, we use parameters corresponding to a commercially available SOA (Thorlabs BOA1554P): saturation power $P_{\text{sat}} = 18$~dBm, maximum small-signal gain $G_{\text{max}} = 35$~dB at maximum current $I_{\text{max}} = 1700$~mA, transparency current $I_{\text{tr}} = 600$~mA, and linewidth enhancement factor $\alpha_H = 5$.

A critical feature of SOAs is the coupling between gain and phase through the linewidth enhancement factor. The SOA transforms the complex field envelope as: \begin{equation} E_{\text{out}} = E_{\text{in}}\,\exp\!\left(\frac{1 - i\alpha_H}{2}\,h\right) \label{eq:soa_field} \end{equation} encapsulating both amplitude gain ($G = e^h$) and a nonlinear phase shift: \begin{equation} \Delta\phi_{\text{SOA}} = -\frac{\alpha_H}{2}\,h \label{eq:phase_shift} \end{equation} This gain--phase coupling means that intensity-dependent gain saturation simultaneously produces intensity-dependent phase modulation, enriching the nonlinear transfer function beyond what either effect alone could provide.

\subsection{Transfer Function}

Each 50/50 directional coupler is described by the unitary transfer matrix: \begin{equation} \mathbf{C} = \frac{1}{\sqrt{2}} \begin{pmatrix} 1 & i \\ i & 1 \end{pmatrix} \label{eq:coupler} \end{equation} For input field $E_0 = \sqrt{P_0}$ entering port~1, the coupler directs $E_0/\sqrt{2}$ to the SOA arm and $iE_0/\sqrt{2}$ to the reference arm. After propagation through the VOA$_1$-SOA-VOA$_2$ cascade (Eqs.~\ref{eq:power_division}--\ref{eq:soa_field}) and the reference-arm phase shifter, the two arm fields are: \begin{align} E_1^{\text{out}} &= \frac{\sqrt{\alpha_1 \alpha_2 G}}{\sqrt{2}}\,e^{-i\alpha_H h/2}\,E_0, \label{eq:arm1_field} \\ E_2^{\text{out}} &= \frac{i}{\sqrt{2}}\,e^{i\phi}\,E_0. \label{eq:arm2_field} \end{align} Applying the output coupler (Eq.~\ref{eq:coupler}) and taking the transmitted port gives the output field: \begin{equation} E_{\text{out}} = \frac{E_0}{2}\left[\sqrt{\alpha_1 \alpha_2 G}\,e^{-i\alpha_H h/2} - e^{i\phi}\right] \label{eq:output_field} \end{equation} The detected output power $P_{\text{out}} = |E_{\text{out}}|^2$ yields the complete module transfer function: \begin{equation} P_{\text{out}} = \frac{P_0}{4}\left[\alpha_1 \alpha_2 G + 1 - 2\sqrt{\alpha_1 \alpha_2 G}\cos\left(\frac{\alpha_H h}{2} + \phi\right)\right] \label{eq:transfer_function} \end{equation} where $G = e^h$ and $h = h(P_{\text{SOA,in}},\, h_0)$ is the solution to Eq.~\eqref{eq:steady_state}. The transfer function has the form of a Mach--Zehnder interference fringe whose visibility, fringe position, and input-dependent curvature are all controlled by the four trainable parameters. The key insight is that $\alpha_1$ affects not only the amplitude scaling but also the degree of gain saturation through Eq.~\eqref{eq:power_division}, coupling the interference condition to the SOA operating point.

\subsection{Role of the two VOAs}

\begin{figure}[H] \centering \includegraphics[width=\textwidth]{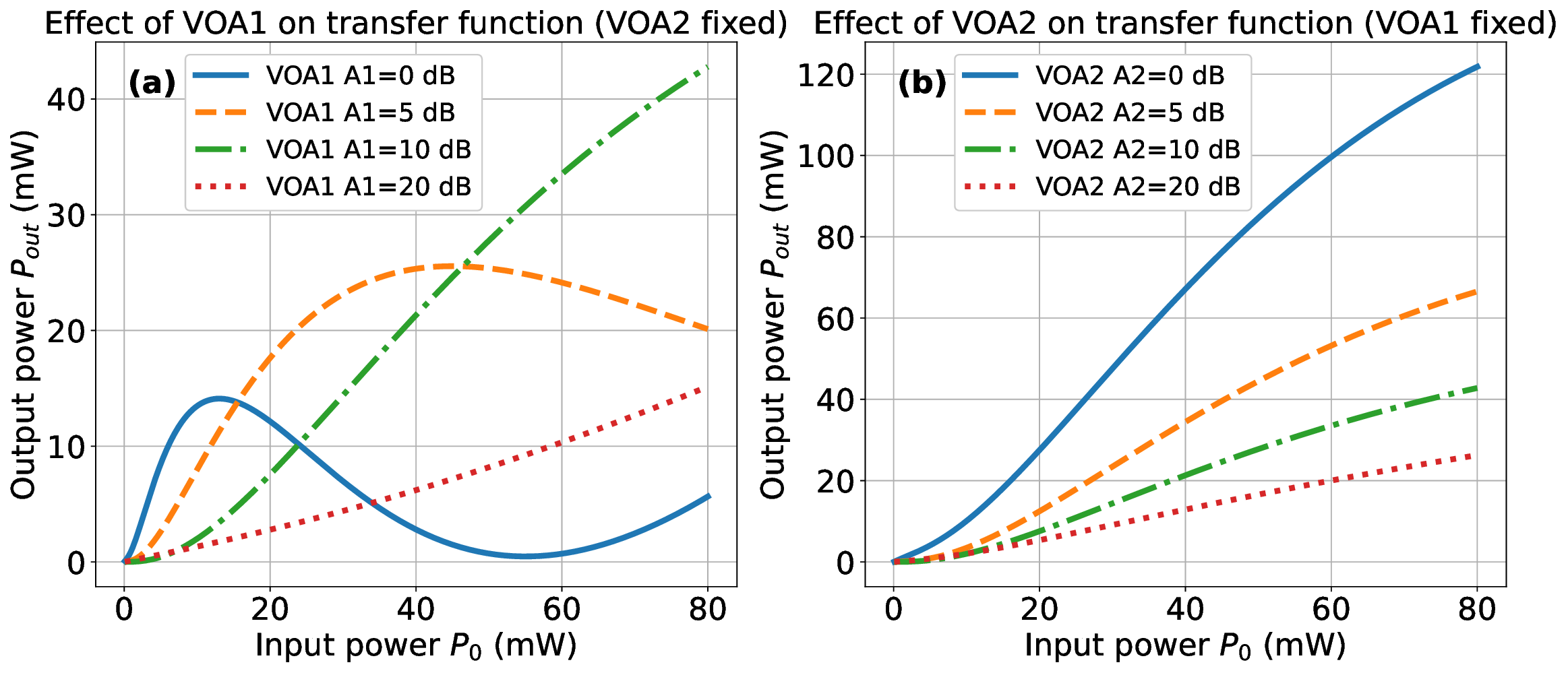} \caption{Distinct roles of the two VOAs in the MZI-VOA-SOA-VOA module transfer function. (a)~Sweeping VOA$_1$ changes the SOA operating point and saturation strength, producing pronounced changes in the nonlinearity shape. (b)~Sweeping VOA$_2$ primarily rescales the SOA-arm contribution at recombination, mainly adjusting interference contrast and overall scaling. Fixed parameters: $I = 1200$~mA, $\phi = \pi/2$, $P_{\mathrm{sat}} = 18$~dBm, $\alpha_H = 5$, with $h_0(I)$ calibrated from $G_{\max} = 35$~dB at $I_{\max} = 1700$~mA and $I_{\mathrm{tr}} = 600$~mA. In~(a) VOA$_2$ is fixed at $A_2 = 10$~dB; in~(b) VOA$_1$ is fixed at $A_1 = 10$~dB.} \label{fig:voa_roles} \end{figure}

The two attenuators provide qualitatively different control knobs over the MZI-VOA-SOA-VOA module. VOA$_1$ (pre-SOA) sets the SOA operating point via $P_{\mathrm{SOA,in}} = \alpha_1 P_0/2$, and therefore tunes the degree of gain saturation and the associated gain-induced phase shift, changing the \emph{shape} of the nonlinear transfer function. In contrast, VOA$_2$ (post-SOA) does not alter the saturation dynamics because it is placed after the SOA; instead, it scales the relative amplitude of the SOA arm at recombination and thus mainly controls the interference contrast and output scaling. Figure~\ref{fig:voa_roles} illustrates these distinct effects by sweeping VOA$_1$ at fixed VOA$_2$ and vice versa, for a representative operating point ($I = 1200$~mA, $\phi = \pi/2$, $P_{\mathrm{sat}} = 18$~dBm, $\alpha_H = 5$, with $h_0(I)$ calibrated from $G_{\max} = 35$~dB at $I_{\max} = 1700$~mA and $I_{\mathrm{tr}} = 600$~mA).

\subsection{Trainable Parameters}

Each MZI-VOA-SOA-VOA module provides four independently trainable parameters. The injection current $I$ (600--1700~mA) controls the small-signal gain $h_0$ through Eq.~\eqref{eq:h0_current}, with higher current producing stronger gain and more pronounced saturation nonlinearity. The input attenuation $\alpha_1$ (0--30~dB) sets the operating point on the saturation curve via Eq.~\eqref{eq:power_division}: high attenuation keeps the SOA in its linear regime, while low attenuation drives it into saturation. The output attenuation $\alpha_2$ (0--30~dB) scales the SOA arm contribution to the interference without affecting the saturation dynamics. Finally, the phase $\phi$ (0--$2\pi$) determines whether the interference is constructive, destructive, or intermediate.

The entire forward pass --- from Eq.~\eqref{eq:power_division} through the Newton--Raphson solution of Eq.~\eqref{eq:steady_state} to the transfer function Eq.~\eqref{eq:transfer_function} --- is implemented in PyTorch with automatic differentiation. Gradients with respect to all four parameters are computed via backpropagation through the Newton iteration, enabling end-to-end training of networks composed of multiple MZI-VOA-SOA-VOA modules.

\subsection{Noise and quantization models}

Practical deployment of photonic neural networks requires robustness to hardware impairments. We model two primary sources of signal degradation: optical noise from amplified spontaneous emission (ASE) in the SOA, and quantization noise from finite-resolution digital-to-analog converters (DACs) used to encode input signals as optical power levels.

\subsubsection{Optical Noise Models}

\cut{SOAs introduce noise through amplified spontaneous emission (ASE), which adds random fluctuations to the optical signal. Our simulation framework implements two noise models with distinct physical assumptions.}

\cut{The amplitude-domain model captures the physics of ASE noise, where random optical fields add coherently to the signal field. For an input power $P_{\text{in}}$, we convert to field amplitude $A = \sqrt{P_{\text{in}}}$, add complex Gaussian noise, and compute the output power:}
\cut{where $n_r$ and $n_i$ are independent random variables drawn from a standard normal distribution and $\sigma = \sqrt{P_{\text{max}}/\text{SNR}_{\text{lin}}}$ sets the noise strength relative to the maximum operating power and the linear-scale signal-to-noise ratio. This model produces signal-dependent variance $\text{Var}[P_{\text{noisy}}] \approx 2 P_{\text{in}} \sigma^2$, meaning low-power signals experience worse relative noise, consistent with the physics of optical detection.}

\add{We model optical noise as additive Gaussian perturbation in the power domain:}
\begin{equation}
P_{\text{noisy}} = P_{\text{in}} + \sigma \cdot n
\label{eq:power_noise}
\end{equation}
\add{where $n \sim \mathcal{N}(0,1)$ and the result is clamped to non-negative values. The noise standard deviation is}
\begin{equation}
\sigma = \sqrt{P_{\text{max}} \cdot 10^{-\text{SNR}_{\text{dB}}/10}}
\label{eq:noise_std}
\end{equation}
\add{where $P_{\text{max}}$ is the maximum operating power. This phenomenological model is deliberately more pessimistic than physical ASE noise: at the operating point ($P_{\text{in}} = 10$--$80$~mW, $G \approx 15$--$30$), the inverse OSNR from amplified spontaneous emission is approximately 0.1--0.3\%, two orders of magnitude below the 14~dB SNR threshold ($\sim$4\% relative noise) used in the robustness analysis. The power-domain model therefore implicitly covers ASE, laser RIN, coupling fluctuations, thermal drift, and detector noise. A full physics-based derivation and four validation experiments are given in Supplementary Section~D.}

\subsubsection{Input Quantization}

The input to each optical module must be encoded as an optical power level, typically using a laser source modulated by a DAC-controlled attenuator or modulator. Finite DAC resolution introduces quantization error. For a $b$-bit DAC spanning the power range $[P_{\text{min}}, P_{\text{max}}]$, the quantized input is obtained by rounding to the nearest discrete level:
\begin{equation}
P_{\text{quant}} = \text{round}\left( \frac{P_{\text{in}} - P_{\text{min}}}{P_{\text{max}} - P_{\text{min}}} \cdot (2^b - 1) \right) \cdot \frac{P_{\text{max}} - P_{\text{min}}}{2^b - 1} + P_{\text{min}}
\label{eq:input_quantization}
\end{equation}
where $2^b - 1$ is the number of discrete levels available (for example, 255 for an 8-bit converter). To enable gradient-based training despite the non-differentiable rounding operation, we employ the straight-through estimator (STE): the forward pass uses quantized values while the backward pass computes gradients as if no quantization occurred.

We evaluate DAC resolutions from 4 to 16 bits. An 8-bit DAC provides 256 levels over the operating range, corresponding to a step size of approximately 0.3~mW for a 10--80~mW range. Our results in Section~\ref{sec2} demonstrate that SSP-KAN maintains high accuracy down to 6-bit resolution, with gradual degradation at lower bit depths.

\subsection{Training procedure}

All SSP-KAN models are trained end-to-end via backpropagation, with optical transfer-function gradients computed by automatic differentiation through the MZI-VOA-SOA-VOA forward model described above. Because no physical hardware is in the loop during training, the procedure reduces to standard gradient-based optimization of the four trainable parameters $(I, \alpha_1, \alpha_2, \varphi)$ per module, with the physics encoded entirely in the differentiable forward pass. \add{All trainable parameters are constrained to the physical operating ranges specified by the device datasheets: injection current $I \in [600,\, 1700]$~mA, attenuations $\alpha_1, \alpha_2 \in [0,\, 30]$~dB, and phase $\phi \in [0,\, 2\pi]$. Each parameter is reparameterised as $\theta = \theta_{\min} + \sigma(z)(\theta_{\max} - \theta_{\min})$, where $z$ is the unconstrained optimiser variable and $\sigma$ is the logistic sigmoid. This smooth bijection ensures that the gradient landscape is smooth and bounded regardless of the raw parameter scale, preventing initialisation-dependent convergence failures. Multi-seed evaluation showed that alternative parameterisations (e.g.\ softplus with post-hoc clamping) left the raw parameter scale uncontrolled, causing certain seeds to converge to narrow minima sensitive to device variation (Supplementary Section~G). The sigmoid reparameterisation eliminates this failure mode. When the differentiable model is unavailable or insufficiently accurate, backpropagation-free training is also feasible: gradient-free optimisers (CMA-ES, PEPG, SPSA) can train SSP-KAN directly from measured network outputs at a cost of a few thousand forward-pass evaluations (Supplementary Section~F). The trained parameter values for every module are reported alongside each result (see Supplementary Information); these values map directly onto experimental set-points.}

\subsubsection{Datasets and preprocessing}

We evaluate SSP-KAN on four benchmarks spanning binary classification, multivariate regression, and high-dimensional image classification (Table~\ref{tab:benchmarks}).

\begin{table}[h]
\caption{Benchmark datasets and preprocessing details.}\label{tab:benchmarks}
\begin{tabular}{@{}lccccc@{}}
\toprule
Dataset & Samples & Input dim. & Task & Split & Power range (mW) \\
\midrule
Two Moons & 2{,}000 & 2 & Classification & 50/50 & 10--80 \\
Yacht Hydro. & 308 & 6 & Regression & 70/15/15 & 10--80 \\
MNIST & 70{,}000 & 784 & Classification & 72/8/20 & 1--10 \\
Fashion-MNIST & 70{,}000 & 784 & Classification & 72/8/20 & 1--10 \\
\botrule
\end{tabular}
\end{table}

For all datasets, inputs are linearly scaled to optical power: $P = P_{\min} + x \cdot (P_{\max} - P_{\min})$, where $x \in [0,1]$ is the normalized input. Two Moons and yacht hydrodynamics use the range 10--80~mW, representative of typical fibre-compatible operating powers. For MNIST and Fashion-MNIST, a power-range sweep identified 1--10~mW as optimal, where the MZI-VOA-SOA-VOA transfer functions operate in a more linear regime better suited to the high-dimensional, low-contrast nature of pixel inputs. The Two Moons dataset comprises 1{,}000 training and 1{,}000 test samples with Gaussian noise ($\sigma = 0.1$), generated using scikit-learn's \texttt{make\_moons}. The yacht hydrodynamics dataset~\cite{UCI,Gerritsma1981} contains 308 samples split 70/15/15 into training, validation, and test sets. MNIST and Fashion-MNIST use the full 70{,}000-sample datasets with stratified 72/8/20 splits, ensuring balanced class representation in each partition.

\subsubsection{Optimization}

All models are trained using the AdamW optimizer~\cite{Loshchilov2019}. For Two Moons and yacht hydrodynamics, which involve small datasets, we use full-batch training for 5{,}000 steps (Two Moons) or 5{,}000 epochs (yacht), with a OneCycleLR schedule (30\% warmup, cosine annealing), weight decay $10^{-5}$, and learning rates of $5 \times 10^{-3}$ (Two Moons) and $10^{-2}$ (yacht SSP-KAN) or $5 \times 10^{-3}$ (yacht baselines). For the yacht task, early stopping with a patience of 500 epochs monitors validation loss to prevent overfitting on the small dataset, and the best-validation-loss checkpoint is restored after training. For MNIST and Fashion-MNIST, we train for 150 epochs with batch size 256, learning rate $10^{-2}$, weight decay $10^{-3}$, and a 5-epoch linear warmup followed by cosine annealing. The best model by validation accuracy is selected. Classification tasks use cross-entropy loss; the regression task uses mean squared error. No data augmentation, label smoothing, or stochastic weight averaging is applied in any experiment.

\subsubsection{Baseline models}

To isolate the contribution of optical nonlinearity, we compare against several baselines trained under identical optimization conditions. A linear model with the same input-output dimensions measures the performance achievable without any nonlinear activation. For Two Moons and yacht hydrodynamics, we additionally compare against software KAN (pyKAN~\cite{Liu2024KAN}) with B-spline grid sizes $G = 1$ and $G = 10$, representing low- and high-capacity software KAN baselines, and a multi-layer perceptron (MLP) with one hidden layer and ReLU activations, sized to approximately match the SSP-KAN parameter count. For image classification, we compare across SSP-KAN architectures of varying depth and width to characterize the effect of network topology.

\subsubsection{Hardware robustness evaluation}

To assess the effect of realistic hardware impairments, we retrain each model from scratch under every combination of input quantization level and optical noise condition, using the same optimizer and schedule as the no-noise baseline. Input quantization is swept from Float32 to 4-bit DAC resolution (16 levels). Optical noise is applied in the power domain (Eq.~\ref{eq:power_noise}) at SNR levels from no noise to 6~dB. This train-under-impairment protocol tests whether the SSP-KAN architecture can learn effective representations when noise and quantization are present throughout training, representing the expected operating condition for physical hardware where impairments are always active.

\add{\subsection{Training and deployment strategies}\label{sec:methods:training_deployment}

The results in this work use offline training: the four per-module parameters are optimised via backpropagation through the differentiable physics model on a standard digital computer. The trained set-points are then transferred to hardware by programming current sources, VOA drivers, and phase controllers directly. This approach assumes that the device-level constants ($P_{\text{sat}}$, $G_{\text{max}}$, $I_{\text{tr}}$) are known from prior characterisation. The primary practical requirement is accurate characterisation of each SOA module. A standard $P_{\text{out}}$ versus $P_{\text{in}}$ sweep at several bias currents suffices to extract the three Agrawal--Olsson parameters. Residual model--device mismatch can be compensated by a fine-tuning pass that re-optimises the four per-module trainable parameters while holding the device constants fixed.

When the physical channel is unknown or the differentiable model is insufficiently accurate, on-site training without backpropagation becomes necessary. Three gradient-free optimisers are suitable for this setting. Covariance Matrix Adaptation Evolution Strategy (CMA-ES)~\cite{Hansen2003} maintains a multivariate Gaussian search distribution over the parameter space and adapts both its mean and covariance at each generation. It offers robust optimisation over multimodal landscapes without gradient information. Parameter-Exploring Policy Gradients (PEPG) estimates the policy gradient by evaluating symmetric perturbations around the current parameter vector, offering lower variance than standard finite-difference methods. Simultaneous Perturbation Stochastic Approximation (SPSA) approximates the gradient using only two function evaluations per iteration regardless of parameter dimensionality. All three operate by evaluating the physical network output directly; no optical backpropagation or adjoint computation is required. Supplementary Section~F presents CMA-ES results on the Two Moons and yacht hydrodynamics tasks.

Thermal drift of the MZI phase is a shared concern for both training strategies. In offline training, drift between characterisation and deployment degrades the fidelity of the transferred set-points. In on-site training, drift during the optimisation process changes the objective landscape between evaluations. For the small-scale configurations studied here (4--7 modules), standard thermo-electric cooler stabilisation and periodic recalibration are expected to be sufficient. Quantitative stability requirements are discussed in Supplementary Section~E.}

\add{\subsection{Energy analysis}\label{sec:methods:energy}

The dominant electrical draw in an SSP-KAN module is the SOA bias current. The BOA1554P datasheet specifies a forward voltage of 1.5~V (typical) to 2.2~V (maximum). At a typical trained operating point ($I \approx 1200$~mA, $V \approx 1.5$~V), each SOA dissipates $P_{\text{elec}} = I \times V \approx 1.8$~W. VOA and phase-shifter biasing are two orders of magnitude below this and are neglected. The [2,2] Two Moons configuration draws $4 \times 1.8 = 7.2$~W across four SOAs; the [6,1,1] yacht configuration draws $7 \times 1.8 = 12.6$~W across seven. At a 1~GHz readout rate, this gives 7.2~nJ and 12.6~nJ per inference. This estimate accounts only for SOA electrical bias; DAC drivers, photodetector electronics, and feedback control are excluded.

For comparison, the equivalent electronic MLP for the yacht task has architecture [6$\rightarrow$4$\rightarrow$1] with 33~parameters. Its arithmetic cost is $6 \times 4 + 4 \times 1 = 28$ multiply-accumulate (MAC) operations per inference. On a 45~nm CMOS process, a single 8-bit MAC costs approximately 1--5~pJ depending on whether operands are served from registers or SRAM~\cite{Horowitz2014,Sze2017}. The resulting 28--140~pJ per inference is a bare-datapath estimate for an idealised ASIC; it excludes clock distribution, control logic, I/O, and data movement beyond the immediate memory hierarchy. System-level energy for a real electronic implementation is higher. SSP-KAN at 7.2~nJ is approximately 50--250$\times$ less energy-efficient than this bare electronic estimate.

The BOA1554P is a fibre-optic booster amplifier designed for high output power in long-haul transmission, with 35~dB gain and 18~dBm saturation power. SSP-KAN does not require these specifications. The architecture requires only sufficient gain saturation to produce a tunable nonlinear transfer function through Eq.~\eqref{eq:transfer_function}. A low-power SOA with $G_{\text{max}} \sim 10$--$15$~dB and lower bias current would draw approximately 50~mW per device. For [2,2] with four such SOAs, the total draw is ${\sim}$200~mW, giving ${\sim}$200~pJ per inference at 1~GHz. This is comparable to the bare electronic MAC estimate (28--140~pJ). The BOA1554P therefore sets an upper bound on per-inference energy; the actual cost scales with device selection.

SSP-KAN does not target raw energy efficiency over electronic ASICs at these scales. The architectural advantages are single-pass latency at group-velocity speed (sub-nanosecond through the module), native optical-domain operation without OEO conversion, and analogue parallelism through wavelength- or space-division multiplexing. A rigorous system-level energy comparison accounting for all overheads on both the optical and electronic sides remains an open benchmarking problem. Device selection and WDM parallelism are the primary levers for reducing per-inference energy.}

\section{Discussion}\label{sec12}
The central question addressed by this work is whether the constrained nonlinear transfer functions of standard telecom components can serve as trainable activations in a structured neural network. Each MZI-VOA-SOA-VOA module provides four tunable parameters --- injection current~$I$, attenuations~$\alpha_1$ and~$\alpha_2$, and phase~$\phi$ --- and the results confirm that this four-parameter family yields sufficient diversity to support nonlinear inference across classification, regression, and image recognition tasks.
\cut{Specifically, SSP-KAN achieves 98.4\% accuracy on nonlinear classification (Two Moons), $R^2 = 0.977$ on six-input multivariate regression (yacht hydrodynamics), and 92.7\% on 784-dimensional image classification (MNIST).} \add{Specifically, SSP-KAN achieves $94.3$\% accuracy (IQR: $90.3$--$97.4$\%) on Two Moons classification ([2,2], 10~seeds), $R^2 = 0.986 \pm 0.015$ on yacht hydrodynamics regression ([6,1,1], 10~seeds), and $93.9 \pm 0.2$\% on MNIST image classification (10~seeds).} All results are obtained from a fully differentiable physics-based model grounded in Agrawal--Olsson SOA gain dynamics~\cite{Agrawal1989} and commercially specified device parameters (Thorlabs BOA1554P), enabling end-to-end optimization via backpropagation.
\cut{The Two Moons and yacht hydrodynamics results illustrate how compositional depth compensates for limited per-module expressivity. Each MZI-VOA-SOA-VOA module implements a cosine-squared nonlinearity modulated by SOA gain saturation (Eq.~\ref{eq:transfer_function}); this four-parameter family is substantially more constrained than B-spline activations with tens or hundreds of degrees of freedom. A single-layer [2,2] SSP-KAN nevertheless achieves 98.4\% accuracy on Two Moons with only four modules, closing to within 1.5~percentage points of the software KAN baseline (pyKAN, 99.9\%). On yacht hydrodynamics, the single-layer [6,1] network ($R^2 = 0.869$) is limited by its inability to capture feature interactions: each module processes one input independently. Adding a second layer ([6,1,1], $R^2 = 0.977$) enables the network to compose six univariate functions, matching pyKAN with fewer than half the parameters (28~vs~60). This confirms that cascading physically constrained nonlinearities produces richer functional representations than increasing parallel width alone. Multi-layer architectures with intermediate optical summation nodes (e.g.\ the [2,2,2] configuration) introduce additional considerations related to coherent phase accumulation, which are analyzed in the Supplementary Material.}
\add{The Two Moons and yacht hydrodynamics results illustrate how compositional depth compensates for limited per-module expressivity. Each MZI-VOA-SOA-VOA module implements a cosine-squared nonlinearity modulated by SOA gain saturation (Eq.~\ref{eq:transfer_function}); this four-parameter family is substantially more constrained than B-spline activations with tens or hundreds of degrees of freedom. The single-layer [2,2] SSP-KAN achieves $94.3$\% (IQR: $90.3$--$97.4$\%) on Two Moons with four modules. The seed variance (best seed: 98.9\%, worst: 88.1\%) indicates multiple local minima in the loss landscape, a consequence of the periodic MZI transfer function. Adding a second layer ([2,2,2], $99.1$\% (IQR: $99.0$--$99.1$\%)) reduces both the accuracy gap and the seed sensitivity. On yacht hydrodynamics, the single-layer [6,1] network ($R^2 = 0.921 \pm 0.017$) is limited by its inability to capture feature interactions: each module processes one input independently. Adding a second layer ([6,1,1], $R^2 = 0.986 \pm 0.015$) enables the network to compose six univariate functions, exceeding the pyKAN baseline ($R^2 = 0.977$) with fewer than half the parameters (28~vs~60). Cascading physically constrained nonlinearities produces richer functional representations than increasing parallel width alone. Multi-layer architectures with intermediate optical summation nodes introduce coherent phase accumulation effects, analyzed in the Supplementary Material.}

\cut{Robustness analyses confirm resilience to the two dominant hardware impairments: finite DAC resolution and optical noise from amplified spontaneous emission. Across both classification and regression tasks, input quantization has negligible impact --- accuracy and $R^2$ remain essentially unchanged from Float32 down to 6-bit resolution under all tested noise conditions. Optical noise is the primary limiting factor. On Two Moons, the [2,2] architecture retains $>$96\% accuracy at SNR~$=$~14~dB; on yacht hydrodynamics, the [6,1,1] network maintains $R^2 > 0.91$ at the same SNR. Under extreme noise (SNR~$=$~6~dB), both tasks exhibit graceful degradation rather than catastrophic failure, with performance converging toward a noise floor that depends on the intrinsic task difficulty rather than on architectural capacity. That quantization effects are negligible while noise effects are measurable indicates that, for practical implementations, investment in low-noise optical amplification will yield greater returns than high-resolution digital-to-analogue conversion.}
\add{Robustness analyses confirm resilience to the two dominant hardware impairments: finite DAC resolution and optical noise from amplified spontaneous emission. Across both tasks, input quantization has negligible impact: accuracy and $R^2$ remain essentially unchanged from Float32 down to 6-bit resolution. Optical noise is the primary limiting factor. On Two Moons, the [2,2] architecture retains ${\sim}92$\% accuracy at 14~dB SNR; on yacht hydrodynamics, the [6,1,1] network maintains $R^2 \approx 0.96$ at the same SNR. Under extreme noise (6~dB), both tasks degrade gracefully, converging toward a noise floor set by intrinsic task difficulty rather than architectural capacity. Quantization effects are negligible while noise effects are measurable; for practical implementations, investment in low-noise optical amplification will yield greater returns than high-resolution digital-to-analogue conversion.}

\add{Physics-based noise analysis confirms that even in a 7-layer cascade, the output SNR remains above 49~dB, more than 35~dB above the phenomenological threshold (Supplementary Section~D). Noise statistics remain near-Gaussian after cascaded nonlinear propagation (skewness $< 0.3$, excess kurtosis $< 0.3$). The phenomenological power-domain model at 14~dB SNR is therefore a conservative bound, approximately $10^2\times$ more pessimistic than the physical ASE noise floor at the CW operating point.}

On MNIST and Fashion-MNIST, SSP-KAN achieves \cut{92.7\% and 85.7\%}\add{$93.9 \pm 0.2$\% and $86.2 \pm 0.2$\%} accuracy, respectively, compared with linear baselines of 91.0\% and 83.3\%. \cut{The modest margin on MNIST reflects the dataset's near-linear separability rather than a limitation of the optical nonlinearity; consistent with this interpretation, the advantage widens on Fashion-MNIST, where nonlinear structure is more pronounced (+2.4 percentage points versus +1.7 on MNIST).}\add{The $+2.9$ percentage point margin over the linear baseline is consistent across both tasks, confirming that the optical nonlinearity provides measurable benefit even on the near-linearly-separable MNIST.} Compared with the dual ring-assisted MZI (D-RAMZI) architecture of Peng \emph{et~al.}~\cite{Peng2024}, which achieves 98\% MNIST accuracy using microring-based nonlinearities with nine tunable parameters per edge and custom silicon photonic fabrication, SSP-KAN employs four parameters per edge using commercially available fibre-coupled components. The D-RAMZI approach offers richer per-edge nonlinearities through resonance-enhanced transmission and cascaded interference; the remaining accuracy gap therefore quantifies the cost of reduced activation richness and suggests that enhancing per-module expressivity represents the most direct route toward closing this gap while retaining deployment simplicity.

It is important to distinguish between architectures used to probe scaling behaviour and those intended for near-term experimental demonstration. The MNIST-scale [784,20,10] network, comprising 15,880 discrete MZI-VOA-SOA-VOA modules, is not proposed as a practical system; assembling such a configuration from fibre-coupled components would be prohibitive in footprint, cost, and thermal stability. Image classification serves instead to characterize SSP-KAN performance trends as input dimensionality and network width increase, informing future integrated implementations. In contrast, the Two Moons and yacht hydrodynamics configurations require only 4--7 modules and are directly realizable on an optical bench using commercially available components.

\add{The scaling studies give initial evidence on how network size relates to performance. On Two Moons, adding a second layer improves accuracy from $94.3$\% ([2,2], 4~modules) to $99.1$\% ([2,2,2], 8~modules) and reduces seed sensitivity. On yacht hydrodynamics, the gain from [6,1] to [6,1,1] is substantial ($R^2$: $0.921 \pm 0.017$ to $0.986 \pm 0.015$), confirming that depth helps most when the task requires feature interactions that a single additive layer cannot capture (Supplementary Section~A). On MNIST, SSP-KAN [784,20,10] reaches $93.9 \pm 0.2$\%, a gain of 2.9 percentage points over the linear baseline. The four-parameter activation family, rather than network capacity, is the limiting factor at scale. Improving SSP-KAN performance on complex tasks will therefore require enhancing per-module expressivity, through cascaded SOA stages, resonant enhancement, or increased parameter count per edge, rather than increasing module count.}

\add{The principal physical limits to scaling are fan-out loss, noise accumulation, and phase stability. At the input layer, each feature is split to $n_{\text{out}}$ edges, so the optical power entering each module scales as $P_{\text{in}} / n_{\text{out}}$. For the [784,20,10] network, each first-layer module receives $P_{\text{in}}/20$, reducing the signal level relative to the SOA's ASE noise floor. At the output of each layer, incoherent summation on the photodetector aggregates contributions from $n_{\text{in}}$ edges, partially recovering the total signal power, but ASE noise from each module accumulates alongside it. The robustness analysis (Sections~\ref{sec:results:moons}--\ref{sec:results:regression}) provides empirical bounds: performance degrades gracefully down to 14~dB SNR, suggesting that cascaded networks can tolerate substantial per-stage noise provided the cumulative SNR remains above this threshold. Phase stability scales linearly with module count, since each MZI requires independent stabilisation. For the 4--7 module configurations that are the experimental targets of this work, the stabilisation overhead is manageable with standard TEC modules and low-bandwidth feedback. Integration onto photonic integrated circuits may become relevant for architectures requiring hundreds of modules, but is beyond the scope of the present work.}

The present study is based on numerical simulations employing calibrated device parameters. Experimental realization is facilitated by the use of components operating well within standard specifications: input powers of 1--80~mW, injection currents of 600--1700~mA, and attenuation ranges of 0--30~dB. \add{Every component in the MZI-VOA-SOA-VOA module is commercially available as a fibre-pigtailed, connectorised device. A near-term prototype can therefore be assembled entirely from catalogue components without custom fabrication or cleanroom access (component specifications and assembly details are given in Supplementary Section~E).} \cut{The primary practical challenges are (i)~accurate characterization of SOA transfer functions under varying bias conditions, and (ii)~maintaining thermal and mechanical phase stability across interferometric modules.} \add{Two primary practical challenges must be addressed for experimental realisation: accurate characterisation of SOA transfer functions under varying bias conditions, and maintaining interferometric phase stability across modules. The Agrawal--Olsson model requires only three measurable device parameters ($P_{\text{sat}}$, $G_{\text{max}}$, $I_{\text{tr}}$), each extractable from a standard characterisation sweep. Active phase stabilisation of fibre-based MZIs is routine in both photonic integrated circuits~\cite{Bogaerts2020,Perez2017} and fibre-optic systems~\cite{Xavier2011}. The SOA power-dependent phase shift operates at MHz--GHz rates, separated from the dither-based stabilisation frequency ($\sim$10~kHz) by more than four decades. It therefore does not corrupt the feedback loop (Supplementary Section~E).} 
The immediate experimental objective is the [2,2] Two Moons configuration, comprising four fibre-coupled modules initialized with parameters obtained from simulation. Successful demonstration would validate both the fidelity of the physics-based model and the end-to-end training framework before scaling to the seven-module yacht hydrodynamics configuration.

\add{A minimal experimental implementation of the [2,2] Two Moons network requires four MZI-VOA-SOA-VOA modules, a tunable laser source, DAC-controlled attenuators for input encoding, and two photodetectors at the output nodes. The training and deployment strategy, including offline and on-site approaches, is described in Methods. The trained parameter values for all modules are reported in Supplementary Section~C.}

\add{Parameter mismatch between trained set-points and their hardware realisations perturbs the transfer function. A systematic device-variability study (Supplementary Section~G) evaluates this sensitivity across four device parameters ($P_{\text{sat}}$, $G_{\text{max}}$, $I_{\text{tr}}$, $\alpha_H$) in 120 training and 6000 Monte Carlo evaluation configurations. Training with variability awareness compensates mismatch on all axes. Without it, $\alpha_H$ dominates: it enters the MZI phase through $\Delta\phi_{\text{SOA}} = -\alpha_H h/2$, so moderate manufacturing variation destroys the learned interference condition. Per-device characterisation of $\alpha_H$ from a standard gain--phase measurement eliminates this vulnerability.}

The SSP-KAN architecture is particularly relevant to signal equalization in coherent optical communications, where inference latency and bandwidth increasingly challenge electronic digital signal processing at next-generation data rates~\cite{Shastri2021}. The per-edge structure maps naturally to wavelength- or space-division multiplexed systems, enabling parallel channel-wise processing. Integration with reservoir computing could combine the temporal feature extraction of SOA-based dynamical systems with the structured nonlinear readout of SSP-KAN. More broadly, the approach of constructing differentiable forward models directly from component physics and optimizing physical parameters via backpropagation extends beyond SOA-based nonlinearities. Any photonic platform with analytically tractable transfer functions (including Kerr-based or Brillouin-based mechanisms) can in principle be incorporated into this physics-aware training paradigm.

\section{Conclusion}\label{sec13}

\cut{We have demonstrated SSP-KAN, a photonic implementation of small-scale Kolmogorov--Arnold Networks in which each network edge is realized by a dedicated MZI--SOA--VOA module acting as a trainable nonlinear activation function. Using only four physically grounded parameters per module --- injection current, two attenuation coefficients, and interferometric phase --- the architecture achieves 98.4\% accuracy on nonlinear classification, $R^2 = 0.977$ on six-input regression, and 92.7\% on 784-dimensional image classification, all through end-to-end differentiable training of SOA gain dynamics using commercially specified device parameters. Compositional depth compensates for the constrained per-module expressivity: cascading physically limited nonlinearities yields richer functional representations than increasing parallel width alone. Robustness analyses confirm that optical noise, rather than DAC resolution, is the dominant hardware impairment, with performance remaining strong down to 6-bit input quantization and SNR~$=$~14~dB across all tasks. Multi-layer architectures with intermediate optical summation introduce coherent phase accumulation effects that are analyzed in the Supplementary Material.}
\add{We have demonstrated SSP-KAN, a photonic implementation of small-scale Kolmogorov--Arnold Networks in which each network edge is realized by a dedicated MZI--SOA--VOA module acting as a trainable nonlinear activation function. Using only four physically grounded parameters per module, the architecture achieves $94.3$\% accuracy (IQR: $90.3$--$97.4$\%) on nonlinear classification ([2,2], 10~seeds), $99.1$\% with a two-layer [2,2,2] network, $R^2 = 0.986 \pm 0.015$ on six-input regression, and $93.9 \pm 0.2$\% on 784-dimensional image classification (10~seeds). Compositional depth compensates for constrained per-module expressivity and reduces seed sensitivity: cascading physically limited nonlinearities yields richer functional representations than increasing parallel width alone. Robustness analyses confirm that optical noise, rather than DAC resolution, is the dominant hardware impairment, with performance remaining above 92\% accuracy on Two Moons and $R^2 > \cut{0.87}\add{0.96}$ on yacht down to 14~dB SNR. Multi-layer architectures with intermediate optical summation introduce coherent phase accumulation effects analyzed in the Supplementary Material.}

The choice between coherent and incoherent source configurations has a depth-dependent effect on both expressivity and trainability. For the single-layer [2,2] network, incoherent power summation (99.4\% accuracy, 16~parameters\add{; single seed, Supplementary Section~B}) substantially outperforms coherent field summation (91.8\%, 20~parameters), despite the latter providing additional phase degrees of freedom: the MZI phase~$\phi$ simultaneously shapes the per-edge transfer function and controls the output interference condition, creating a multimodal loss landscape that impedes gradient-based training. For the two-layer [2,2,2] network, this balance reverses: coherent operation (99.9\%, 40~parameters) slightly exceeds the incoherent baseline (99.5\%, 32~parameters), as the interference cross-terms at intermediate nodes introduce input-dependent effective weights that enrich the network's functional capacity. This depth-dependent crossover indicates that incoherent summation is preferable for shallow, experimentally accessible configurations, while coherent architectures may become advantageous as network depth increases and alternative optimization strategies are employed.

Clear limitations remain. The restricted four-parameter activation family produces a measurable accuracy gap relative to architectures employing resonant or highly engineered nonlinearities, such as the D-RAMZI design of Peng \emph{et~al.}~\cite{Peng2024}. This gap quantifies the cost of prioritizing experimental accessibility over per-edge expressivity, and indicates that enhancing activation richness (through cascaded stages, resonant enhancement, or alternative mechanisms such as stimulated Brillouin scattering or Kerr nonlinearity) represents the most direct route to closing it while retaining telecom compatibility. Furthermore, while the small-scale configurations studied here (4--7 modules) are directly realizable on an optical bench, scaling to high-dimensional tasks will require transition to integrated photonic platforms for compactness, thermal stability, and cost.

The fully differentiable physics model developed here provides a general framework for co-designing photonic neural network architectures with their physical substrates: any optical platform with analytically tractable transfer functions can in principle be incorporated into this optimization paradigm. The immediate experimental objective is the four-module [2,2] Two Moons configuration, initialized with simulation-trained parameters, to validate both model fidelity and the end-to-end training framework before scaling to the seven-module yacht hydrodynamics system. Beyond component-level validation, the per-edge SSP-KAN structure maps naturally to wavelength- or space-division multiplexed fibre systems, positioning it for latency-critical applications such as optical channel equalization where electronic digital signal processing increasingly struggles to keep pace with data rates.

\backmatter

\bmhead{Supplementary information}
Supplementary Information is available for this paper. It includes \cut{two} \add{seven} appendices: architectural limitations of single-layer SSP-KAN (Supplementary Section~A, Supplementary Fig.~1)\cut{and}, coherent versus incoherent operation analysis (Supplementary Section~B, Supplementary Fig.~2)\add{, trained module parameters for all experimentally realisable configurations (Supplementary Section~C, Supplementary Tables~1--4), physics-based noise model derivation and validation (Supplementary Section~D, Supplementary Tables~5--6, Supplementary Figs.~3--5), experimental realisation details (Supplementary Section~E), gradient-free training results (Supplementary Section~F), and device parameter variability analysis (Supplementary Section~G, Supplementary Tables~8--10)}. Three supplementary videos show the training evolution of learned decision boundaries on Two Moons for the [2,2] incoherent (Supplementary Video~1), [2,2,2] incoherent (Supplementary Video~2), and [2,2,2] coherent (Supplementary Video~3) architectures.

\bmhead{Acknowledgements}

LNC acknowledges the support by the Engineering and Physical Sciences Research Council (EPSRC) under grant EP/Z534614/1, as part of the European Training Network on Post-Digital Computing+ (POSTDIGITAL+) funded through the Marie Sk\l{}odowska-Curie Actions Horizon Europe programme, Europe Grant Agreement 101169118 HORIZON-MSCA-2023-DN-01-01. SKT acknowledges the support of the EPSRC project EP/W002868/1. EM acknowledges the support of the Royal Society (IF\textbackslash R2\textbackslash 25 2078)
SKT and EM acknowledge the support of the UK Multidisciplinary Centre for Neuromorphic Computing (UKRI982).

\section*{Declarations}

\begin{itemize}
\item \textbf{Funding:} Engineering and Physical Sciences Research Council (EPSRC), grant EP/Z534614/1 (POSTDIGITAL+, Marie Sk\l{}odowska-Curie Actions Horizon Europe); EPSRC, grant EP/W002868/1; Royal Society, grant IF\textbackslash R2\textbackslash 25 2078; UK Multidisciplinary Centre for Neuromorphic Computing (UKRI982).
\item \textbf{Conflict of interest:} The authors declare no competing interests.
\end{itemize}

\section*{Data availability}

The Two Moons dataset is generated synthetically using \texttt{sklearn.datasets.make\_moons}. The UCI Yacht Hydrodynamics dataset is publicly available~\cite{UCI}. MNIST and Fashion-MNIST are standard benchmarks available through PyTorch (\texttt{torchvision.datasets}). Source Data are provided with this paper.

\section*{Code availability}

Source code for SSP-KAN, including the differentiable MZI-VOA-SOA-VOA model and all experiment scripts, will be made available on GitHub upon publication.

\bibliography{sn-bibliography}

\clearpage

\begin{appendices}

\section*{Supplementary Information: SSP-KAN --- Photonic Kolmogorov--Arnold Networks Using Saturated Semiconductor Optical Amplifiers}

\section{Architectural limitations of single-layer SSP-KAN} \label{sec:supp:limitations}

\begin{figure}[H] \centering \includegraphics[width=\textwidth]{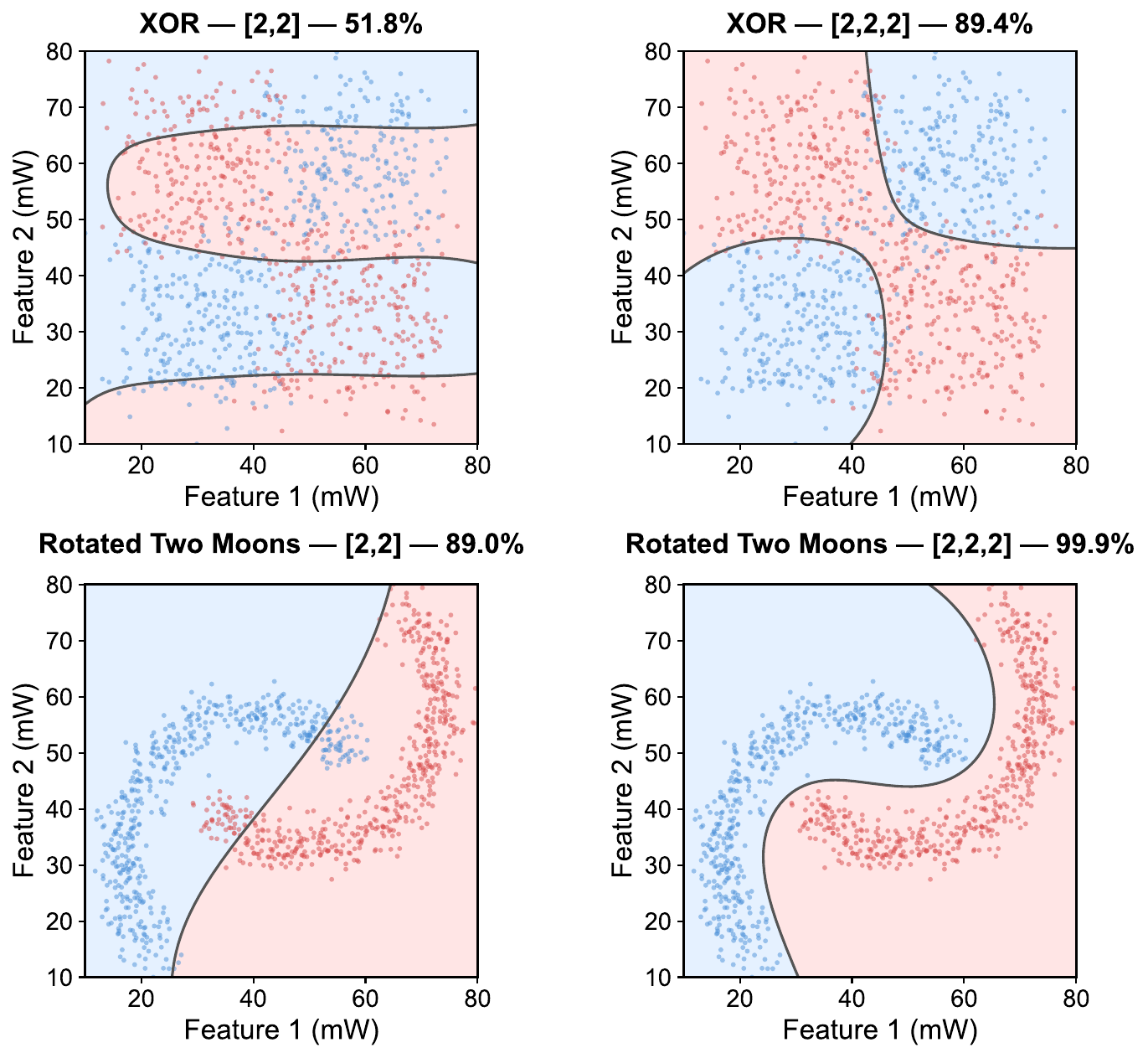} \caption{Tasks requiring feature interaction. Left column: single-layer [2,2]; right column: two-layer [2,2,2]. Top row: XOR. The [2,2] boundary is at chance (51.8\%) while [2,2,2] recovers 89.4\%. Bottom row: Two Moons rotated 45\textdegree. The [2,2] network (89.0\%) falls to the linear baseline because the additive structure cannot represent diagonal boundaries, while [2,2,2] reaches 99.9\%.} \label{fig:supp_limitations} \end{figure}

A single-layer [2,2] SSP-KAN computes each output as a sum of univariate functions, $y_j = \varphi_{1,j}(x_1) + \varphi_{2,j}(x_2)$. This additive structure cannot represent decision boundaries that depend on interactions between inputs. Two tasks illustrate the limitation (Supplementary Fig.~\ref{fig:supp_limitations}).

The XOR task assigns opposite labels to diagonally opposed quadrants, requiring a boundary that depends jointly on both features. The [2,2] network achieves only 51.8\% accuracy, indistinguishable from random chance, because no sum of univariate functions can partition the plane into four quadrants. Adding a second layer recovers 89.4\%: the [2,2,2] network composes first-layer outputs to approximate the diagonal structure, though the task remains challenging owing to the sharp axis-aligned boundaries that smooth optical transfer functions approximate imperfectly.

A subtler failure arises when the Two Moons dataset is rotated by 45\textdegree. In its standard orientation the crescent boundary is approximately separable along the coordinate axes, and the [2,2] network reaches \cut{98.4\%}\add{$94.3$\% (IQR: $90.3$--$97.4$\%, 10~seeds)} (main text, Section~2.1). After rotation the boundary runs diagonally, breaking this separability: the [2,2] network drops to 89.0\%, close to the linear baseline. The [2,2,2] network restores \cut{99.9\%}\add{$99.1$\%} by composing first-layer features to reconstruct the rotated separation. This pair of results, \cut{98.4\%}\add{$94.3$\%} at 0\textdegree{} versus 89.0\% at 45\textdegree{}, \cut{directly }quantifies the cost of the additive constraint and motivates multi-layer architectures, which in turn raise the coherent phase accumulation effects analyzed in Supplementary Section~\ref{sec:supp:coherent}.

\section{Coherent versus incoherent operation} \label{sec:supp:coherent}

Throughout the main text, module outputs combine as optical powers, which is valid when the recombining fields are mutually incoherent. If instead all paths originate from a single laser, the fields retain a fixed phase relationship and combine coherently, producing interference terms that alter the network computation. This section derives the incoherent and coherent summation regimes and compares them on the Two Moons benchmark.

At each output or intermediate node~$j$, the $n_\text{in}$ modules routed to that node contribute fields that recombine through a coupler tree. In the most general form, the detected power is \begin{equation} y_j = \biggl|\frac{1}{\sqrt{n_\text{in}}} \sum_{i=1}^{n_\text{in}} c_{ij}\,E_{f_{ij},\text{out}} \biggr|^2, \label{eq:supp_coherent_general} \end{equation} where $c_{ij}$ is the phase factor accumulated by the field traveling from input~$i$ to output~$j$ through the coupler network. Expanding the square produces $n_\text{in}$ direct terms and $n_\text{in}(n_\text{in}{-}1)/2$ cross-terms, each depending on the relative phase between a pair of modules. The two regimes correspond to whether these cross-terms survive or vanish.

\subsection{Incoherent regime}

When the combining fields originate from separate lasers, or when path-length mismatch exceeds the coherence length, the cross-terms average to zero and only powers add. For a [2,2] layer the outputs reduce to \begin{align} y_1 &= \tfrac{1}{2}\bigl[ f_1(x_1 P_0/4) + f_3(x_2 P_0/4)\bigr], \label{eq:supp_incoh_y1} \\ y_2 &= \tfrac{1}{2}\bigl[ f_2(x_1 P_0/4) + f_4(x_2 P_0/4)\bigr], \label{eq:supp_incoh_y2} \end{align} where $f_k$ is the power-domain transfer function of module~$k$ and the factors $1/4$ and $1/2$ arise from successive 50/50 couplers. The effective weight on every edge is fixed at~$1/2$; only the four module parameters $(I, \alpha_1, \alpha_2, \phi)$ per edge are trainable. This is the regime assumed throughout the main text (16 trainable parameters for [2,2]).

\subsection{Coherent regime}

When all signals share the same source, the coupler phase factors $c_{ij}$ in Eq.~\eqref{eq:supp_coherent_general} must be tracked. Each lossless 50/50 coupler follows the transfer matrix: \begin{equation} \mathbf{C} = \frac{1}{\sqrt{2}} \begin{pmatrix} 1 & i \\ i & 1 \end{pmatrix}, \label{eq:supp_coupler} \end{equation} where the bar port transmits unchanged and the cross port acquires a factor~$i$. A field traversing $n_\times$ cross ports through a cascade of couplers therefore accumulates a phase factor $i^{n_\times}$.

For a [2,2] layer fed by a coherent source $E_0 = \sqrt{P_0}$, inputs are encoded via amplitude modulators ($E \to \sqrt{x}\,E$) and each arm is split once more to address two modules. Each field traverses two couplers: module~1 takes two bar ports (phase factor $i^0 = +1$), modules~2 and~3 each take one cross port ($i^1 = +i$), and module~4 takes two cross ports ($i^2 = -1$). In our simulation we label each path by a binary index $k$ encoding bar~(0) or cross~(1) at each stage, so that the phase factor reduces to $i^{\,\mathrm{popcount}(k)}$.

The resulting input fields are \begin{align} E_{f_1,\text{in}} &= +\frac{\sqrt{x_1}\,E_0}{2}, &\qquad E_{f_2,\text{in}} &= +\frac{i\sqrt{x_1}\,E_0}{2}, \label{eq:supp_fields_upper}\\[4pt] E_{f_3,\text{in}} &= +\frac{i\sqrt{x_2}\,E_0}{2}, &\qquad E_{f_4,\text{in}} &= -\frac{\sqrt{x_2}\,E_0}{2}. \label{eq:supp_fields_lower} \end{align} The sign pattern $(+1,\,+i,\,+i,\,-1)$ is fixed by coupler unitarity and enters every downstream interference term.

Each module acts on the complex field as \begin{equation} E_{f_k,\text{out}} = t_k\!\bigl(|E_{f_k,\text{in}}|^2\bigr)\; e^{\,i\theta_k(|E_{f_k,\text{in}}|^2)} \cdot E_{f_k,\text{in}}, \label{eq:supp_field_transfer} \end{equation} where $t_k$ is the amplitude transmission and $\theta_k$ the total output phase, both set by the SOA gain dynamics and the module parameters. Expanding Eq.~\eqref{eq:supp_coherent_general} for the [2,2] case (modules~1 and~3 feeding output~1) gives \begin{equation} y_1 = \frac{1}{2}\bigl[ |E_1|^2 + |E_3|^2 + 2|E_1||E_3|\sin(\theta_1 - \theta_3) \bigr]. \label{eq:supp_interference} \end{equation} The cross-term has two consequences. First, because $\theta_k$ depends on input power through SOA gain saturation, the effective weights at recombination nodes vary with the signal. Second, each module's MZI phase $\phi_k$ now controls both the transfer function shape and the interference condition, coupling two roles that are independent in the power-domain formulation.

A phase shifter $\psi_k$ placed after each module separates these roles: \begin{equation} y_1 = \frac{1}{2}\bigl[|E_1|^2 + |E_3|^2 + 2|E_1||E_3|\sin(\theta_1 - \theta_3 + \Delta\psi_{13})\bigr], \label{eq:supp_with_psi} \end{equation} where $\Delta\psi_{13} = \psi_1 - \psi_3$ controls the interference independently of the MZI-VOA-SOA-VOA parameters that shape the activation. This adds one parameter per module (5~per edge versus~4 in the incoherent regime).

\subsection{Experimental comparison}

\begin{figure}[H] \centering \includegraphics[width=\textwidth]{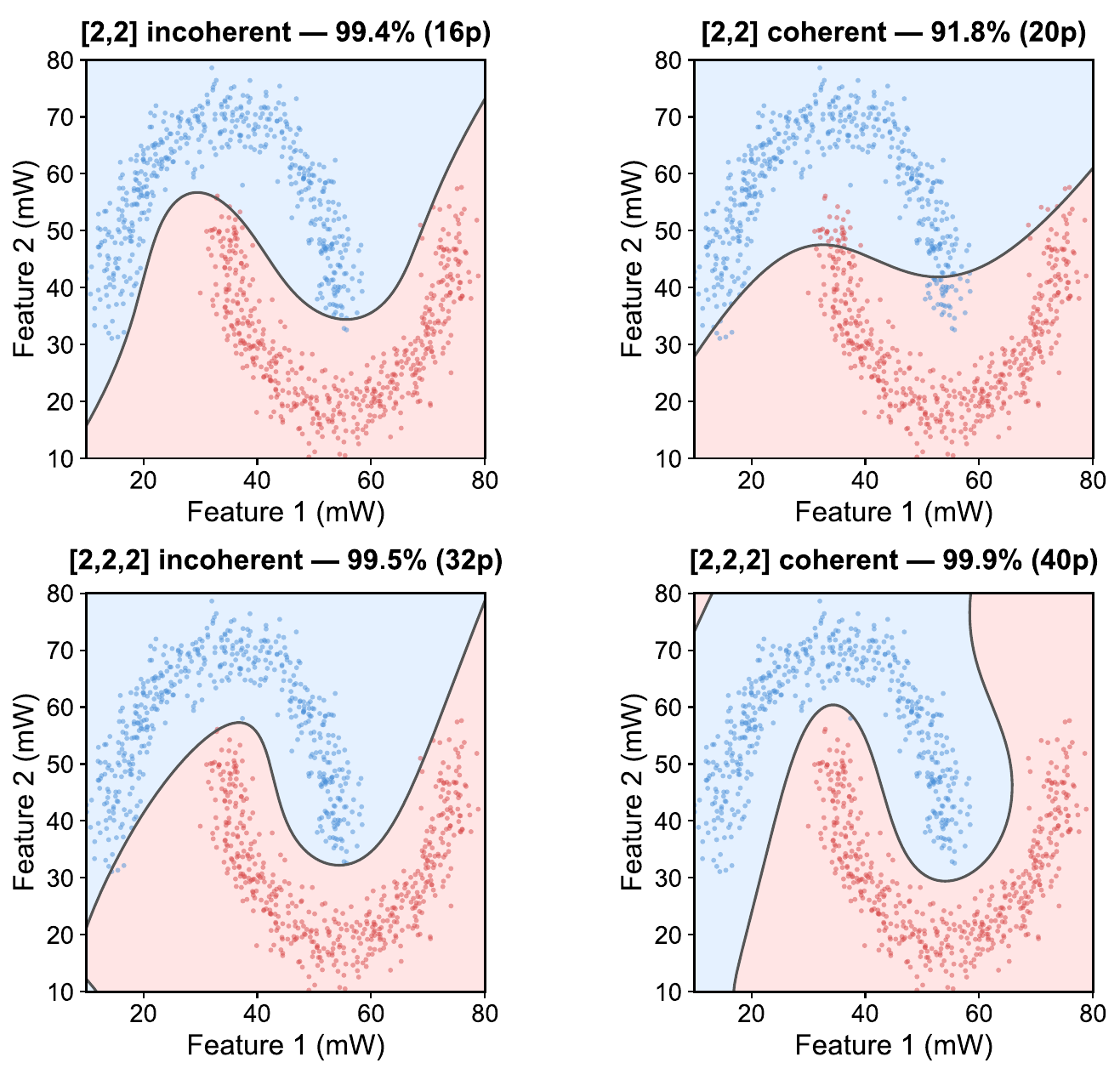} \caption{Coherent versus incoherent operation on Two Moons classification\add{ (single seed)}. Top row: single-layer [2,2]. The incoherent regime (99.4\%, 16 parameters) outperforms the coherent regime (91.8\%, 20 parameters), indicating that interference at the output couplers complicates optimization without providing a net benefit at this depth. Bottom row: two-layer [2,2,2]. The coherent regime (99.9\%, 40 parameters) slightly exceeds the incoherent regime (99.5\%, 32 parameters), as the interference cross-terms at intermediate nodes introduce input-dependent effective weights that increase network expressivity. \add{Multi-seed statistics for the incoherent regime are reported in the main text.}} \label{fig:supp_coherent} \end{figure}

Supplementary Fig.~\ref{fig:supp_coherent} compares both regimes on Two Moons classification for [2,2] and [2,2,2] architectures.

For the single-layer [2,2] network, the incoherent regime (99.4\%, 16 parameters\add{; single seed --- multi-seed statistics in Section~2.1 of the main text}) substantially outperforms the coherent regime (91.8\%, 20 parameters). Although coherent operation provides the additional $\psi$ degrees of freedom, it also couples $\phi_k$ to the interference condition at the output couplers. Since the MZI transfer function already depends on $\phi_k$ through a cosine-squared fringe, this coupling multiplies the number of local minima in the loss landscape: the same parameter now sits inside two periodic functions with different periods. Gradient descent is more likely to become trapped, and in this case settles on an inferior boundary. The performance gap reflects the difficulty of gradient-based optimization on the resulting multimodal landscape, not a limitation of the coherent model, which is strictly more expressive. Alternative training strategies which handle multimodal objectives more naturally may recover the missing performance.

For the two-layer [2,2,2] network, the situation reverses: the coherent regime (99.9\%, 40 parameters) slightly exceeds the incoherent regime (99.5\%, 32 parameters). Here the interference cross-terms at intermediate nodes act as input-dependent effective weights that the incoherent formulation cannot access. The additional $\psi$ parameters allow the optimizer to exploit this extra expressivity without the $\phi$-interference coupling penalty dominating.

The choice of source coherence therefore affects both expressivity and trainability, and the balance between them depends on network depth. For the small-scale configurations studied in the main text, the incoherent regime offers a simpler optimization landscape at no cost in performance for [2,2] and [6,1,1]. The coherent regime becomes advantageous only when intermediate recombination nodes are present and the network is deep enough to exploit the additional interference degrees of freedom.

\section{Trained module parameters} \label{sec:supp:params}

All trainable parameters are clamped to the physical operating ranges specified by the Thorlabs BOA1554P datasheet: injection current $I \in [600,\, 1700]$~mA, attenuations $\alpha_1, \alpha_2 \in [0,\, 30]$~dB, and phase $\phi \in [0,\, 2\pi]$~rad. The values reported below are the effective parameters after the softplus and clamp transformations applied during the forward pass, and correspond directly to the hardware set-points required for experimental realisation.

\subsection{Two Moons classification}

\begin{table}[H]
\caption{Trained parameters for the [2,2] SSP-KAN on Two Moons (4 modules, 16 parameters\cut{, test accuracy 98.4\%}\add{; mean test accuracy $94.3$\% (IQR: $90.3$--$97.4$\%) across 10~seeds; values shown are for a representative seed}).}\label{tab:params_moons_22}
\centering
\begin{tabular}{@{}lcccc@{}}
\toprule
Module & $I$ (mA) & $\alpha_1$ (dB) & $\alpha_2$ (dB) & $\phi$ (rad) \\
\midrule
$\varphi_{0,0}$ & 1045.26 & 0.54 & 0.28 & 1.4926 \\
$\varphi_{0,1}$ & 1363.19 & 0.70 & 3.69 & 3.2287 \\
$\varphi_{1,0}$ & 1136.59 & 3.09 & 0.07 & 1.4761 \\
$\varphi_{1,1}$ & 1219.03 & 0.28 & 7.91 & 1.8061 \\
\botrule
\end{tabular}
\end{table}

\begin{table}[H]
\caption{Trained parameters for the [2,2,2] SSP-KAN on Two Moons (8 modules, 32 parameters\cut{, test accuracy 99.2\%}\add{; mean test accuracy $99.1$\% (IQR: $99.0$--$99.1$\%) across 10~seeds; values shown are for a representative seed}).}\label{tab:params_moons_222}
\centering
\begin{tabular}{@{}llcccc@{}}
\toprule
Layer & Module & $I$ (mA) & $\alpha_1$ (dB) & $\alpha_2$ (dB) & $\phi$ (rad) \\
\midrule
1 & $\varphi_{0,0}$ & 1035.54 & 2.17 & 2.15 & 4.6215 \\
1 & $\varphi_{0,1}$ & 1358.99 & 0.02 & 0.22 & 3.3808 \\
1 & $\varphi_{1,0}$ & 1117.07 & 3.25 & 1.67 & 2.1629 \\
1 & $\varphi_{1,1}$ & 1222.74 & 6.14 & 4.19 & 4.6445 \\
\midrule
2 & $\varphi_{0,0}$ & 931.23 & 2.53 & 0.25 & 4.5404 \\
2 & $\varphi_{0,1}$ & 951.79 & 1.97 & 4.62 & 2.5132 \\
2 & $\varphi_{1,0}$ & 833.05 & 0.96 & 0.11 & 5.1438 \\
2 & $\varphi_{1,1}$ & 908.22 & 1.81 & 0.12 & 1.3007 \\
\botrule
\end{tabular}
\end{table}

\subsection{Yacht hydrodynamics regression}

\begin{table}[H]
\caption{Trained parameters for the [6,1] SSP-KAN on yacht hydrodynamics (6 modules, 24 parameters\cut{, test $R^2 = 0.869$}\add{; mean test $R^2 = 0.921 \pm 0.017$ across 10~seeds; values shown are for a representative seed}).}\label{tab:params_yacht_61}
\centering
\begin{tabular}{@{}lcccc@{}}
\toprule
Module & $I$ (mA) & $\alpha_1$ (dB) & $\alpha_2$ (dB) & $\phi$ (rad) \\
\midrule
$\varphi_{0,0}$ & 955.69 & 10.50 & 0.09 & 0.4920 \\
$\varphi_{1,0}$ & 1169.22 & 15.78 & 1.13 & 3.2527 \\
$\varphi_{2,0}$ & 950.79 & 10.64 & 0.11 & 0.7227 \\
$\varphi_{3,0}$ & 1056.65 & 13.26 & 0.56 & 5.0791 \\
$\varphi_{4,0}$ & 971.82 & 11.24 & 0.07 & 0.2775 \\
$\varphi_{5,0}$ & 1308.35 & 3.07 & 7.73 & 3.2592 \\
\botrule
\end{tabular}
\end{table}

\begin{table}[H]
\caption{Trained parameters for the [6,1,1] SSP-KAN on yacht hydrodynamics (7 modules, 28 parameters\cut{, test $R^2 = 0.977$}\add{; mean test $R^2 = 0.986 \pm 0.015$ across 10~seeds; values shown are for a representative seed}).}\label{tab:params_yacht_611}
\centering
\begin{tabular}{@{}llcccc@{}}
\toprule
Layer & Module & $I$ (mA) & $\alpha_1$ (dB) & $\alpha_2$ (dB) & $\phi$ (rad) \\
\midrule
1 & $\varphi_{0,0}$ & 960.90 & 6.73 & 3.18 & 0.8875 \\
1 & $\varphi_{1,0}$ & 1176.99 & 7.65 & 7.22 & 4.7838 \\
1 & $\varphi_{2,0}$ & 956.79 & 3.84 & 8.26 & 2.2538 \\
1 & $\varphi_{3,0}$ & 1060.58 & 8.77 & 5.24 & 5.9785 \\
1 & $\varphi_{4,0}$ & 977.85 & 6.16 & 6.10 & 1.1401 \\
1 & $\varphi_{5,0}$ & 1300.58 & 2.62 & 2.18 & 3.6593 \\
\midrule
2 & $\varphi_{0,0}$ & 930.61 & 0.36 & 0.05 & 2.9707 \\
\botrule
\end{tabular}
\end{table}


\section{Physics-based noise model derivation and validation} \label{sec:supp:noise}

This section derives physics-based noise models for the MZI-VOA-SOA-VOA module, validates them against the manufacturer-specified noise figure of the Thorlabs BOA1554P, and compares their predictions against the phenomenological noise models used in the main text.

\subsection{Current-dependent ASE model} \label{sec:supp:noise:current_dependent}

\subsubsection{Spontaneous emission factor}

The spontaneous emission factor $n_{\text{sp}}$ quantifies the quality of population inversion in the gain medium. For a semiconductor optical amplifier with linear material gain $g = a(N - N_0)$, where $a$ is the differential gain, $N$ is the carrier density, and $N_0$ is the transparency carrier density~\cite{Henry1986}:
\begin{equation}
n_{\text{sp}} = \frac{N}{N - N_0}
\label{eq:supp_nsp}
\end{equation}
At high inversion ($N \gg N_0$), $n_{\text{sp}} \to 1$. Near transparency ($N \to N_0^+$), $n_{\text{sp}} \to \infty$. The practical range for InGaAsP SOAs is 1.4--4~\cite{Henry1986}.

\subsubsection{Connection to the trainable parameter $I$}

The carrier density obeys the rate equation~\cite{Baveja2010}:
\begin{equation}
\frac{dN}{dt} = \frac{I}{qV} - R(N) - \Gamma v_g\, g(N)\, S
\label{eq:supp_rate}
\end{equation}
where $I$ is the injection current, $q$ is the electron charge, $V$ is the active region volume, $R(N)$ is the total recombination rate, $\Gamma$ is the optical confinement factor, $v_g$ is the group velocity, and $S$ is the photon density. In the unsaturated regime ($P_{\text{in}} \ll P_{\text{sat}}$), the stimulated emission term is negligible and $N \propto I$. At transparency, $N = N_0$ at $I = I_{\text{tr}}$. Substituting into Eq.~\eqref{eq:supp_nsp}:
\begin{equation}
\boxed{n_{\text{sp,unsat}} = \frac{I/I_{\text{tr}}}{I/I_{\text{tr}} - 1}}
\label{eq:supp_nsp_unsat}
\end{equation}
At $I = 1200$~mA: $n_{\text{sp,unsat}} = 2.0$. At $I = 1700$~mA: $n_{\text{sp,unsat}} \approx 1.55$.

\subsubsection{Saturation correction}

At the SSP-KAN operating point ($P_{\text{in}}/P_{\text{sat}} \approx 0.16$--$1.27$), the stimulated emission term in Eq.~\eqref{eq:supp_rate} dominates. Carriers are depleted by the strong optical signal, clamping $N$ below the unsaturated value and pushing it toward $N_0$. This increases $n_{\text{sp}}$.

Using the integrated gain $h = \Gamma a (N - N_0) L$, the saturated and unsaturated cases give:
\begin{equation}
N_{\text{sat}} - N_0 = \frac{h}{h_0}(N_{\text{unsat}} - N_0)
\label{eq:supp_nsat}
\end{equation}
Substituting into $n_{\text{sp}} = N_{\text{sat}}/(N_{\text{sat}} - N_0)$:
\begin{equation}
\boxed{n_{\text{sp,sat}} = 1 + \frac{h_0}{h}\,(n_{\text{sp,unsat}} - 1)}
\label{eq:supp_nsp_sat}
\end{equation}
At no saturation ($h = h_0$): $n_{\text{sp,sat}} = n_{\text{sp,unsat}}$. At deep saturation ($h \ll h_0$): $n_{\text{sp}}$ grows as $h_0/h$, reflecting the degradation of inversion quality. Both $h$ and $h_0$ are already computed in the forward pass (Newton--Raphson solver and Baveja model respectively), so this correction adds no new parameters.

\subsubsection{ASE spectral density}

The single-sided ASE power spectral density per polarisation mode at the SOA output is~\cite{Desurvire1994}:
\begin{equation}
S_{\text{ASE}} = n_{\text{sp}}(G - 1)\,h\nu \quad [\text{W/Hz, per polarisation mode}]
\label{eq:supp_sase}
\end{equation}
where $G = e^h$ is the saturated power gain and $h\nu = 1.28 \times 10^{-19}$~J at 1550~nm.

\subsubsection{ASE propagation through one MZI-VOA-SOA-VOA module}

The MZI has two arms recombined at a 50:50 output coupler. ASE is broadband incoherent, so at the coupler it adds as power, not field. Tracing the ASE through each path:

\begin{center}
\begin{tabular}{@{}ll@{}}
\toprule
Path & ASE contribution at output \\
\midrule
Active arm (existing ASE) & $S_{\text{in}} \cdot \alpha_1 \alpha_2 G / 4$ \\
Active arm (new ASE from SOA) & $\alpha_2 \cdot n_{\text{sp}}(G-1)h\nu / 2$ \\
Reference arm (existing ASE) & $S_{\text{in}} / 4$ \\
\botrule
\end{tabular}
\end{center}

Summing:
\begin{equation}
S_{\text{out}} = \frac{1}{4}(\alpha_1 \alpha_2 G + 1)\,S_{\text{in}} + \frac{1}{2}\,\alpha_2\,n_{\text{sp}}(G-1)\,h\nu
\label{eq:supp_sout}
\end{equation}
where $S_{\text{in}}$ is the incoming ASE spectral density from the previous layer (zero for the first layer).

\subsubsection{Multi-layer ASE accumulation}

At each output node $j$ in a KAN layer, ASE from multiple edges sums as power (broadband incoherent):
\begin{equation}
S_{\text{ASE},j} = \sum_i S_{\text{ASE,out},ij}
\label{eq:supp_sase_sum}
\end{equation}
Signal power $P_j$ and ASE spectral density $S_{\text{ASE},j}$ are tracked as two separate quantities through the network. In a multi-layer network (e.g.\ [2,2,2]), layer~2's SOAs receive layer~1's accumulated ASE as $S_{\text{in}}$ in Eq.~\eqref{eq:supp_sout}. The $\alpha_1 \alpha_2 G$ factor in the first term represents re-amplification of upstream ASE by downstream SOAs. However, the total ASE power ($P_{\text{ASE}} = S_{\text{ASE}} \times B_o \approx 0.05~\mu$W) remains five orders of magnitude below the signal power ($P_{\text{in}} = 10$--$80$~mW), so ASE does not saturate downstream SOAs.

\subsubsection{Detection noise at the photodetector}

Detection noise is computed once, at the final photodetector. Three noise terms arise from square-law detection~\cite{Personick1973,Mukai1982}:

\paragraph{Signal--ASE beat (dominant).} The coherent signal field and the broadband ASE field beat against each other at the photodetector:
\begin{equation}
\sigma^2_{\text{sig-sp}} = 4\,P_{\text{signal}}\,S_{\text{ASE}}\,B_e \quad [\text{W}^2]
\label{eq:supp_sigsp}
\end{equation}
where $S_{\text{ASE}}$ is per polarisation mode. The factor~4 arises from the heterodyne geometry: 2 from mixing with ASE on both sides of the signal frequency, and 2 from both quadratures of the beat~\cite{Mukai1982}.

\paragraph{Shot noise (negligible at high gain).}
$\sigma^2_{\text{shot}} = 2\,h\nu\,P_{\text{signal}}\,B_e$. The ratio $\sigma^2_{\text{shot}}/\sigma^2_{\text{sig-sp}} \sim 1/(2 n_{\text{sp}} G)$, which is $\sim 10^{-3}$ at our operating point ($\sim$35~dB below signal--ASE beat).

\paragraph{ASE--ASE beat (negligible for strong signals).}
$\sigma^2_{\text{sp-sp}} = 4\,S_{\text{ASE}}^2\,(2 B_o B_e - B_e^2)$, where $B_o$ is the optical filter bandwidth ($\sim$1--5~THz). This term dominates only when the signal is absent or very weak; for our regime ($P_{\text{signal}} \gg P_{\text{ASE}}$), it is negligible compared to signal--ASE beat.

\subsubsection{CW validity bound}

The gain equation used throughout this work sets $dh/dt = 0$, assuming steady-state carrier density. This requires the observation time $\Delta t$ to be much longer than the carrier lifetime $\tau_c$:
\begin{equation}
\Delta t \gg \tau_c \quad \Rightarrow \quad B_e \approx \frac{1}{\Delta t} \ll \frac{1}{\tau_c}
\label{eq:supp_cw_bound}
\end{equation}
For the BOA1554P with $\tau_c \approx 200$--$500$~ps (typical bulk InGaAsP): $1/\tau_c \approx 2$--$5$~GHz. Self-consistency therefore requires $B_e \lesssim 1$~GHz. Since $\sigma^2_{\text{sig-sp}} \propto B_e$ (Eq.~\eqref{eq:supp_sigsp}), the noise is bounded to be small in exactly the regime where the CW model is valid.

\subsection{Datasheet noise model} \label{sec:supp:noise:ds}

The BOA1554P datasheet specifies a fibre-to-fibre NF of 6--8~dB, corresponding to a linear noise figure:
\begin{equation}
F = 10^{\text{NF}_{\text{dB}}/10} \approx 4\text{--}6.3
\label{eq:supp_F}
\end{equation}
For a high-gain phase-insensitive amplifier, $F \approx 2\,n_{\text{sp}}$~\cite{Henry1986}, giving:
\begin{equation}
n_{\text{sp}} \approx F/2 \approx 2\text{--}3.15
\label{eq:supp_nsp_ds}
\end{equation}
This measured NF already encodes all internal physics: spatially varying carrier density, internal waveguide loss, and fibre coupling loss. No approximation chain is needed. The same propagation equations (Eqs.~\eqref{eq:supp_sout}--\eqref{eq:supp_sigsp}) apply, with $n_{\text{sp}}$ fixed regardless of operating point.

The datasheet validates the \emph{unsaturated} limit of the current-dependent model. At small signal ($h \to h_0$), Eq.~\eqref{eq:supp_nsp_sat} reduces to $n_{\text{sp,unsat}}$, and $F \approx 2\,n_{\text{sp,unsat}}$ matches the measured NF. The saturation correction itself (the $h_0/h$ factor) has not been validated against input-power-dependent NF measurements for the BOA1554P.

\subsection{Validation} \label{sec:supp:noise:exp}

To compare the four noise models under identical training conditions, we train SSP-KAN under nine noise configurations (baseline at 30~dB, power-domain at 14/20/30~dB, amplitude-domain at 14/20/30~dB, datasheet NF at 8~dB, and current-dependent) across three architectures with 10 random seeds each. Supplementary Table~\ref{tab:supp_exp1} reports the results side by side.

\begin{table}[H]
\caption{Noise model comparison across architectures and tasks (10~seeds, mean $\pm$ std). Both physics-based models produce performance statistically indistinguishable from the 30~dB baseline.}\label{tab:supp_exp1}
\centering
\begin{tabular}{@{}lccc@{}}
\toprule
Noise model & Moons [2,2] (\%) & Moons [2,2,2] (\%) & Yacht [6,1,1] $R^2$ \\
\midrule
Baseline (30~dB) & $95.2 \pm 3.6$ & $98.0 \pm 3.3$ & $0.977 \pm 0.011$ \\
Power 14~dB & $90.6 \pm 2.5$ & $91.5 \pm 4.0$ & $0.945 \pm 0.014$ \\
Power 20~dB & $91.5 \pm 3.0$ & $94.8 \pm 3.4$ & $0.967 \pm 0.012$ \\
Power 30~dB & $92.2 \pm 3.4$ & $97.3 \pm 3.3$ & $0.975 \pm 0.012$ \\
Amplitude 14~dB & $73.9 \pm 0.3$ & $64.0 \pm 0.5$ & $0.121 \pm 0.173$ \\
Amplitude 20~dB & $84.0 \pm 0.5$ & $75.9 \pm 0.3$ & $0.462 \pm 0.127$ \\
Amplitude 30~dB & $90.5 \pm 2.6$ & $86.8 \pm 2.7$ & $0.818 \pm 0.144$ \\
Datasheet NF 8~dB & $92.2 \pm 3.4$ & $97.8 \pm 3.3$ & $0.977 \pm 0.011$ \\
Current-dependent & $92.0 \pm 3.4$ & $97.9 \pm 3.3$ & $0.977 \pm 0.011$ \\
\botrule
\end{tabular}
\end{table}

The two physics-based models (datasheet NF and current-dependent) are indistinguishable from each other and from the baseline across all tasks, confirming that physical ASE noise at the CW operating point is negligible for inference. The amplitude-domain model is substantially worse at every SNR: at 14~dB, yacht $R^2$ drops to $0.121 \pm 0.173$, compared with $0.945 \pm 0.014$ for the power-domain model at the same nominal SNR. The difference arises because the amplitude model applies noise proportional to the signal field, so high-power signals receive disproportionately more noise. The power-domain model distributes noise uniformly regardless of signal level. Since the SSP-KAN operating range (10--80~mW) places the signal well above the noise floor, the amplitude model penalises strong signals far more than weak ones, collapsing the effective dynamic range.

\begin{figure}[H] \centering
\includegraphics[width=\textwidth]{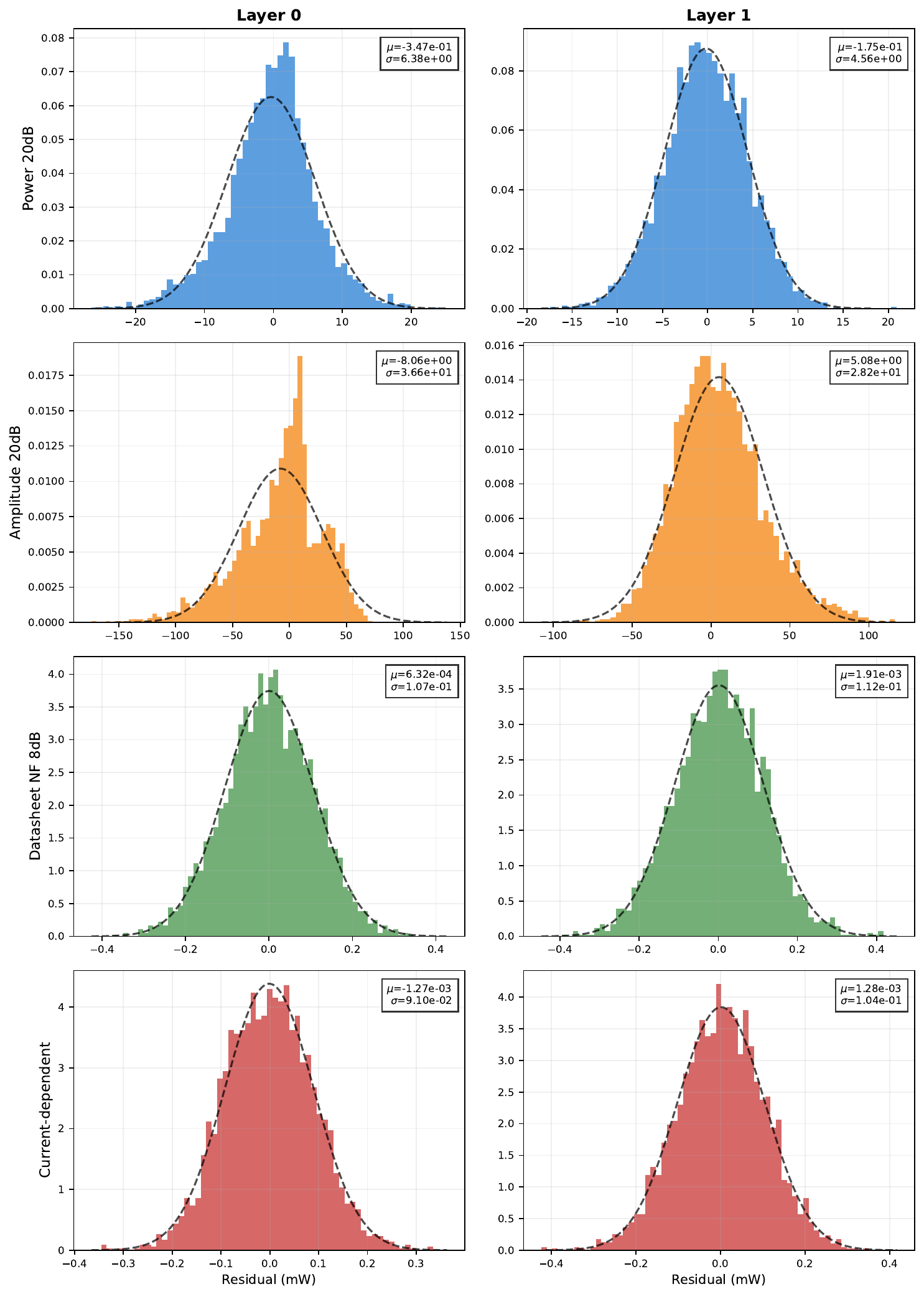}
\caption{Noise residual distributions (noisy output $-$ baseline output) at the output of the [2,2,2] network for four noise models (rows) and two layers (columns), pooled across three representative inputs. The physics-based models (datasheet NF and current-dependent) produce residuals with $\sigma < 0.1$~mW, approximately $10^4\times$ smaller than the phenomenological models. Dashed lines show Gaussian fits.}
\label{fig:supp_histograms}
\end{figure}

A concern for SOA-based networks is whether the nonlinear transfer function distorts the noise statistics. If gain saturation introduced heavy tails or skewness, the Gaussian assumptions underlying the phenomenological models would break down at physical noise levels. To test this, we propagate 1000 Two Moons test samples through the trained [2,2,2] network and record the noise residuals (noisy output minus baseline output) at each layer (Supplementary Fig.~\ref{fig:supp_histograms}). For both physics-based models, residuals after layer~1 remain near-Gaussian: skewness $< 0.3$ and excess kurtosis $< 0.3$ for all input classes. The phenomenological models at higher noise levels produce larger residuals but similar distributional shape. Non-Gaussian distortion from SOA saturation is negligible at physical noise levels.

\begin{figure}[H] \centering
\includegraphics[width=0.55\textwidth]{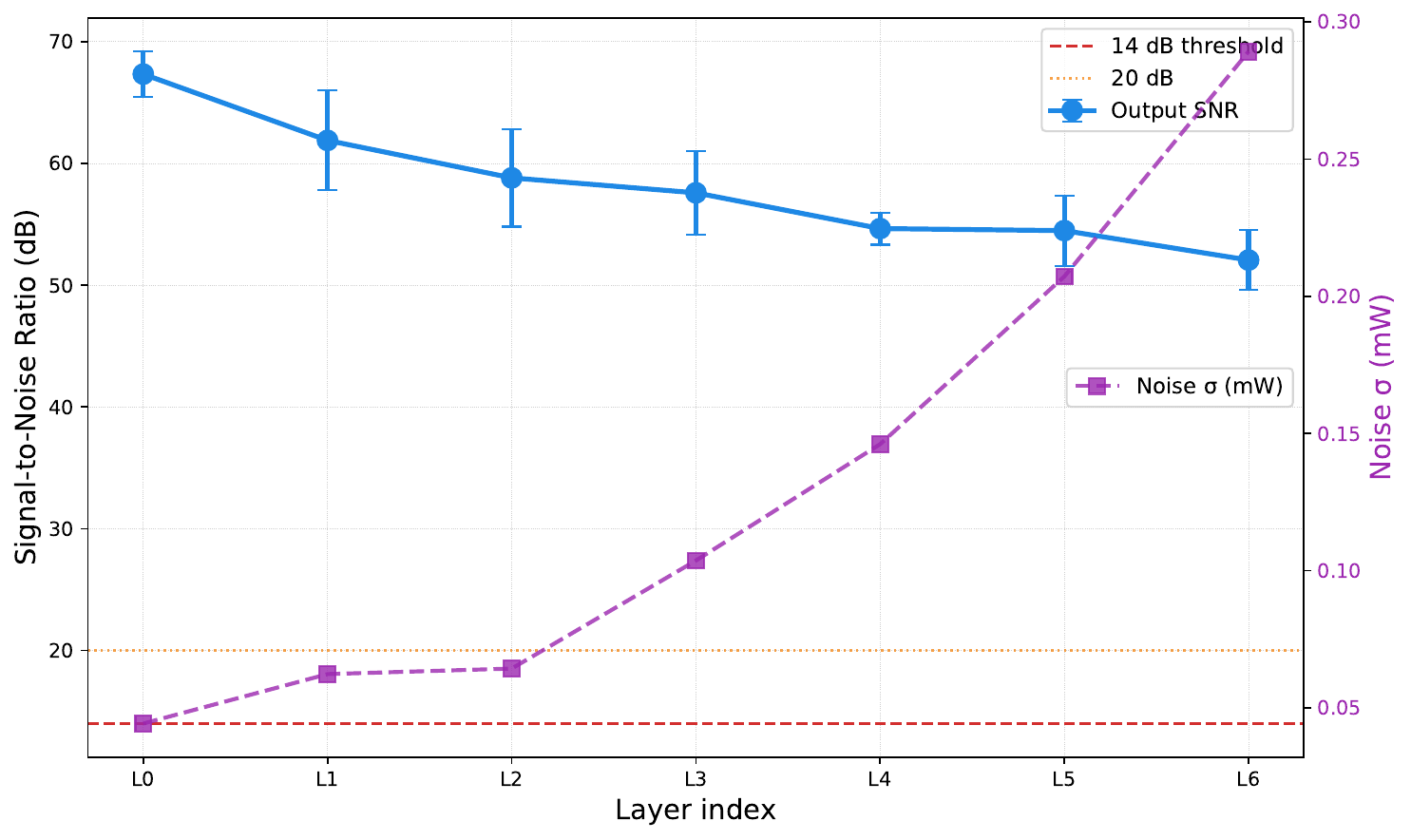}
\caption{SNR at each layer output through a 7-layer [2,2,\ldots,2] cascade under current-dependent ASE noise (10~seeds, mean $\pm$ std). The dashed line marks the 14~dB phenomenological threshold from the main text. SNR decreases from ${\sim}$67~dB to ${\sim}$49~dB across seven layers but remains above the threshold by more than 35~dB.}
\label{fig:supp_cascade}
\end{figure}

The main text identifies noise accumulation through cascaded layers as a scaling concern: ASE from first-layer SOAs is re-amplified by downstream SOAs, and the signal--ASE beat noise compounds at each detection stage. To quantify this beyond the architectures used in the main text, we train a 7-layer [2,2,\ldots,2] network on Two Moons and measure the SNR at each layer output under current-dependent ASE noise (Supplementary Fig.~\ref{fig:supp_cascade}). SNR decreases monotonically from ${\sim}$67~dB at layer~0 to ${\sim}$49~dB at layer~6. Each layer adds ASE, but re-amplification maintains the signal well above the noise floor. At depth~7 the accumulated SNR remains above the 14~dB phenomenological threshold by more than 35~dB.

\begin{figure}[H] \centering
\includegraphics[width=0.7\textwidth]{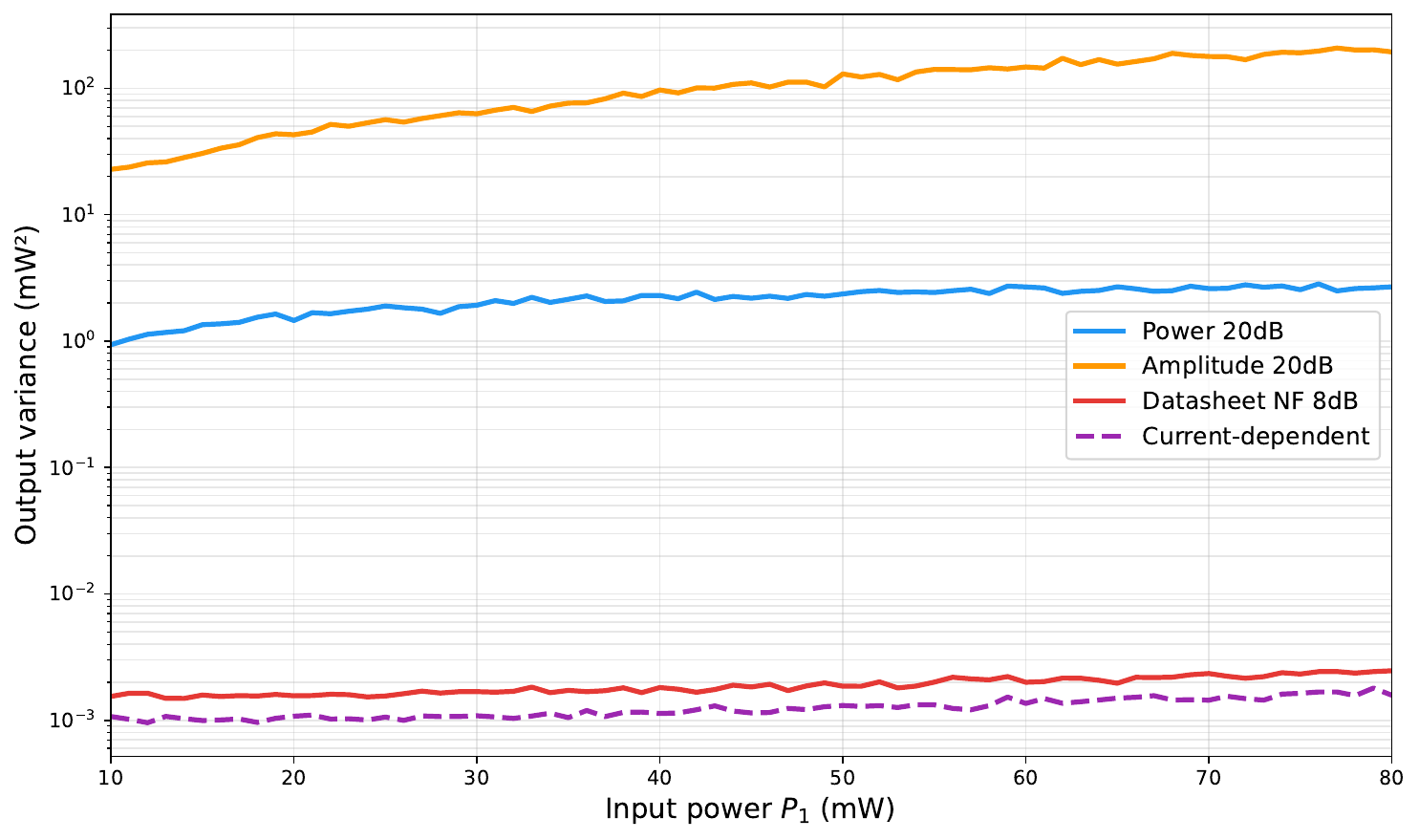}
\caption{Output variance versus input power for four noise models. The physics-based models (bottom curves) show variance proportional to signal power, consistent with signal--ASE beat noise. Log-scale $y$-axis.}
\label{fig:supp_variance}
\end{figure}

The amplitude-domain model in Supplementary Table~\ref{tab:supp_exp1} shows the strongest signal dependence: noise power scales with signal power, so high-power signals are penalised disproportionately. The same dependence appears in the physics-based models through the signal--ASE beat term ($\sigma^2_{\text{sig-sp}} \propto P_{\text{signal}}$, Eq.~\eqref{eq:supp_sigsp}). To quantify this, we sweep input power from 10 to 80~mW through a trained [2,2] module and record output variance over 1000 noise realisations per model (Supplementary Table~\ref{tab:supp_variance}, Supplementary Fig.~\ref{fig:supp_variance}). The physics-based models produce output variance approximately $1.4 \times 10^4$ times smaller than the phenomenological power-domain model. Both exhibit $\sigma^2 \propto P_{\text{signal}}$, confirming the heterodyne beat mechanism, but the maximum output standard deviation under current-dependent noise is ${\sim}$0.03~mW, compared with signal powers of 10--80~mW.

\begin{table}[H]
\caption{Output variance range across input power for each noise model. The physics-based models produce variance approximately $10^4\times$ smaller than the phenomenological power-domain model.}\label{tab:supp_variance}
\centering
\begin{tabular}{@{}lcccc@{}}
\toprule
Noise model & Min var.\ [mW$^2$] & Max var.\ [mW$^2$] & Ratio & Signal dependence \\
\midrule
Power 20~dB & $4.16 \times 10^{-1}$ & $5.46 \times 10^{-1}$ & 1.3 & Weak \\
Amplitude 20~dB & $1.70 \times 10^{1}$ & $2.52 \times 10^{1}$ & 1.5 & Moderate \\
Datasheet NF 8~dB & $1.13 \times 10^{-3}$ & $1.78 \times 10^{-3}$ & 1.6 & $\sigma^2 \propto P_{\text{signal}}$ \\
Current-dependent & $8.89 \times 10^{-4}$ & $1.15 \times 10^{-3}$ & 1.3 & $\sigma^2 \propto P_{\text{signal}}$ \\
\botrule
\end{tabular}
\end{table}

In summary, physical ASE noise is negligible at the CW operating point. Even in a 7-layer cascade, the SNR remains above 49~dB, more than 35~dB above the phenomenological threshold where performance degrades. The phenomenological power-domain model at 14~dB SNR is approximately $10^2\times$ more pessimistic than the physical ASE floor and is a conservative stress test covering impairment sources beyond ASE alone.


\section{Experimental realisation details} \label{sec:supp:experimental}

This section details the practical considerations for experimental realisation of SSP-KAN: SOA characterisation procedures, interferometric phase stability, and the minimal experimental configuration.

\subsection{SOA characterisation and model calibration}

The Agrawal--Olsson model used in the main text (Eq.~(5) of the main text) is parameterised by three device-level quantities: saturation power $P_{\text{sat}}$, maximum small-signal gain $G_{\text{max}}$ (equivalently $h_{0,\text{max}}$), and transparency current $I_{\text{tr}}$. For the Thorlabs BOA1554P, the datasheet specifies $G_{\text{max}} \approx 27$~dB (small-signal, fibre-to-fibre), saturation output power $P_{\text{sat,out}} = 18.5$~dBm, and operating current range 600--1700~mA~\cite{BOA1554P}. The simulations in the main text use $G_{\text{max}} = 35$~dB, $P_{\text{sat}} = 18$~dBm, and $I_{\text{tr}} = 600$~mA; these values represent the gain within the waveguide (before fibre coupling loss) and should be refined from measured data.

\paragraph{Extraction procedure.} Each SOA module is characterised by measuring output power $P_{\text{out}}$ as a function of input power $P_{\text{in}}$ at several injection currents spanning 600--1700~mA. The transparency current $I_{\text{tr}}$ is the bias at which fibre-to-fibre transmission crosses 0~dB. The small-signal gain $G_{\text{max}}(I)$ is obtained from $P_{\text{out}}/P_{\text{in}}$ in the low-power limit ($P_{\text{in}} \ll P_{\text{sat}}$) at each current. The saturation power $P_{\text{sat}}$ follows from the 3~dB compression point: the Agrawal--Olsson model gives $P_{\text{sat}} \approx P_{3\text{dB}}/\ln 2$. The gain coefficient $\kappa = h_{0,\text{max}}/(I_{\text{max}}/I_{\text{tr}} - 1)$ then determines the current-to-gain mapping (Eq.~(10) of the main text).

\paragraph{Linewidth enhancement factor.} The variability study (Supplementary Section~\ref{sec:supp:variability}) identifies $\alpha_H$ as the single most sensitive device parameter. Extracting $\alpha_H$ per device is therefore critical for deployment. The standard method modulates the SOA injection current at a frequency $\omega$ and measures the resulting amplitude modulation (AM) and phase modulation (PM) on the output signal simultaneously. The ratio of PM to AM indices gives $\alpha_H$ directly: $\alpha_H = -2\,(\Delta\phi / \Delta G)$, where $\Delta\phi$ is the phase modulation depth and $\Delta G$ is the fractional gain modulation. This measurement requires only a coherent receiver or a fibre interferometer and is routinely performed for SOA and laser characterisation. For the BOA1554P, $\alpha_H$ is expected to lie in the range 3--7 depending on wavelength and bias current, consistent with the sampling range used in the variability study.

\paragraph{Fine-tuning protocol.} Once all five device-level constants ($P_{\text{sat}}$, $G_{\text{max}}$, $I_{\text{tr}}$, $\alpha_H$, and the gain coefficient $\kappa$) are measured, the four per-module trainable parameters ($I$, $\alpha_1$, $\alpha_2$, $\phi$) can be re-optimised via backpropagation through the differentiable model with the measured constants substituted. This fine-tuning pass compensates for residual model--device mismatch (e.g.\ from wavelength dependence, temperature sensitivity, or the linear-gain approximation) without requiring a full retraining of the network.

\subsection{Phase stability}

\cut{Interferometric phase stability is a known challenge for fibre-based MZI systems. The standard mitigation techniques used in coherent optical communications and fibre sensing apply directly to the 4--7 module configurations targeted here: thermo-electric cooler (TEC) stabilisation of each MZI assembly, low-bandwidth feedback locking with a pilot tone or dither signal, and path-matched design to minimise arm-length imbalance. Stabilisation overhead scales linearly with module count; for hundreds of modules, photonic integration becomes necessary.}

\add{SSP-KAN requires active phase stabilisation. The MZI phase $\phi$ is a trained parameter; any drift from the trained set-point distorts the transfer function and degrades inference accuracy.}

\add{Active MZI stabilisation is a mature technology in both integrated and fibre-optic platforms. Programmable photonic circuits based on MZI meshes routinely stabilise tens to hundreds of interferometers using on-chip thermal tuners and local feedback loops~\cite{Bogaerts2020,Perez2017}. SSP-KAN requires stabilising only 4--7 MZIs, far fewer than typical programmable processors. For fibre-pigtailed implementations, piezoelectric fibre stretchers driven by PID controllers achieve sub-radian phase stability at kHz servo bandwidths, as demonstrated over 1~km fibre paths~\cite{Xavier2011}. The bench-scale SSP-KAN modules present a less demanding stabilisation target.}

\add{A concern specific to SSP-KAN is that the SOA introduces a power-dependent phase shift $\Delta\phi_{\text{SOA}} = -\alpha_H h / 2$ that varies with the input signal. This could in principle corrupt a dither-based stabilisation loop. The relevant timescales are, however, separated by more than four orders of magnitude. Environmental drift (thermal, mechanical) operates at 1--100~Hz. The dither frequency for phase stabilisation is typically 1--10~kHz. The signal-dependent phase modulation follows the input at MHz to GHz rates, bounded by the CW validity condition $B_e \ll 1/\tau_c \approx 2$--$5$~GHz (Eq.~\ref{eq:supp_cw_bound}). Because the stabilisation loop integrates over many signal periods, the power-dependent phase shift averages to its mean value and does not corrupt the feedback signal. The loop tracks only the slowly varying geometric phase offset between the two MZI arms.}

\add{Stabilisation overhead scales linearly with module count. For the 4--7 module configurations targeted here, standard TEC modules and low-bandwidth feedback are sufficient. Transition to photonic integrated circuits becomes relevant for architectures requiring hundreds of modules.}

\subsection{Minimal experimental setup}

The [2,2] Two Moons classifier requires a tunable CW laser source ($P_{\text{out}} \geq 20$~mW, C-band), two DAC-controlled variable optical attenuators for input encoding (0--30~dB, $\geq$8-bit resolution), two 1:2 fibre-optic couplers for input fan-out, four MZI-VOA-SOA-VOA modules, two 2:1 couplers for output summation, and two InGaAs photodetectors at the output nodes. Each module comprises two 50/50 directional couplers, one SOA (Thorlabs BOA1554P or equivalent), two variable optical attenuators, and one fibre stretcher or thermo-optic phase shifter for the MZI phase $\phi$. Four TEC modules (one per MZI) provide phase stabilisation, and a digital control interface (FPGA or microcontroller with DAC outputs) programs the current sources and VOA drivers.

All components are commercially available as fibre-pigtailed, connectorised devices. The total component cost is estimated at \$5,000--10,000, dominated by the four SOAs at approximately \$1,000 each. No cleanroom access, custom photonic chip fabrication, or free-space optical alignment is required.


\section{Gradient-free training results} \label{sec:supp:cmaes}

When the differentiable physics model is insufficiently accurate or the physical channel is unknown, gradient-free optimisation can train SSP-KAN directly from measured network outputs. We evaluate Covariance Matrix Adaptation Evolution Strategy (CMA-ES)~\cite{Hansen2003} as a representative gradient-free optimiser.

\subsection{Experimental setup}

CMA-ES maintains a multivariate Gaussian search distribution over the parameter space (16 dimensions for [2,2], 28 for [6,1,1]) and adapts both its mean and covariance at each generation. We use initial step size $\sigma_0 = 0.3$, population size $\lambda = 20$ (Moons) or $\lambda = 24$ (yacht), and run for 200 generations (Moons) or 300 generations (yacht). Each generation requires $\lambda$ forward-pass evaluations of the physical network. The total evaluation budget is therefore 4,000 (Moons) and 7,200 (yacht).

\subsection{Results}

\begin{table}[H]
\caption{CMA-ES versus Adam+backpropagation. CMA-ES achieves comparable mean performance with substantially lower variance across seeds.}\label{tab:supp_cmaes}
\centering
\begin{tabular}{@{}llcccc@{}}
\toprule
Task & Optimiser & Mean metric & Std & Best & Worst \\
\midrule
Moons [2,2] & Adam (500 epochs) & 94.2\% & 1.0\% & 95.8\% & 92.4\% \\
Moons [2,2] & CMA-ES (4,000 evals) & 93.8\% & 0.0\% & 93.8\% & 93.8\% \\
\midrule
Yacht [6,1,1] & Adam (1,000 epochs) & $R^2 = 0.881$ & 0.198 & 0.983 & 0.447 \\
Yacht [6,1,1] & CMA-ES (7,200 evals) & $R^2 = 0.955$ & 0.048 & 0.986 & 0.879 \\
\botrule
\end{tabular}
\end{table}

On Two Moons, CMA-ES achieves 93.8\% accuracy with zero variance across 5~seeds, compared with Adam's 94.2\% $\pm$ 1.0\%. The 0.4~percentage point difference in mean accuracy is offset by CMA-ES's complete elimination of seed-dependent variability.

On yacht hydrodynamics, CMA-ES achieves $R^2 = 0.955 \pm 0.048$, substantially outperforming Adam's $R^2 = 0.881 \pm 0.198$ in both mean and worst-case performance. Adam's high variance (worst seed: $R^2 = 0.447$) indicates sensitivity to initialisation in this higher-dimensional parameter space (28 parameters); CMA-ES's covariance adaptation provides more robust convergence.

These results demonstrate that SSP-KAN can be trained without backpropagation at a cost of 4,000--7,200 forward-pass evaluations. For a physical network operating at CW with microsecond-scale measurement times, this corresponds to total training times on the order of seconds to minutes, making on-site training practical for laboratory deployment.


\section{Device parameter variability} \label{sec:supp:variability}

Each MZI-VOA-SOA-VOA module depends on four device-level constants that vary across manufactured devices: saturation power $P_{\text{sat}}$, maximum small-signal gain $G_{\text{max}}$, transparency current $I_{\text{tr}}$, and linewidth enhancement factor $\alpha_H$. Training uses nominal datasheet values (Supplementary Table~\ref{tab:supp_variability_params}). A deployed device will differ from these values. This section quantifies the sensitivity of inference accuracy to device-to-device variation on the yacht [6,1,1] regression task and identifies which parameters require per-device characterisation.

\begin{table}[H]
\caption{Device parameter sampling ranges. Nominal values correspond to the Thorlabs BOA1554P parameters used throughout the main text.}\label{tab:supp_variability_params}
\centering
\begin{tabular}{@{}lccc@{}}
\toprule
Parameter & Nominal & Sampling range & Units \\
\midrule
$G_{\text{max}}$ & 35.0 & $\mathcal{U}[32.0,\, 35.0]$ & dB \\
$P_{\text{sat}}$ & 18.0 & $\mathcal{U}[17.0,\, 19.0]$ & dBm \\
$I_{\text{tr}}$ & 600 & $\mathcal{U}[\text{nom}\!\cdot\!(1{-}f),\; \text{nom}\!\cdot\!(1{+}f)]$ & mA \\
$\alpha_H$ & 5.0 & $\mathcal{U}[3.0,\, 7.0]$ & --- \\
\botrule
\end{tabular}
\end{table}

Eight perturbation conditions are tested: nominal (no perturbation), five single-parameter ablations ($G_{\text{max}}$, $P_{\text{sat}}$, $\alpha_H$, $I_{\text{tr}}$ at ${\pm}\,5$\%, and $I_{\text{tr}}$ at ${\pm}\,10$\%), and two combined conditions (all four parameters varied simultaneously, $I_{\text{tr}}$ spread $f = 5$\% or $10$\%). Two deployment scenarios are evaluated. In the first, each module's device constants are sampled at model initialisation and the optimiser trains with them in place, simulating deployment with a known variability profile. In the second, the model is trained on nominal parameters and then evaluated under 50 Monte Carlo device draws per condition, simulating direct transfer to an uncharacterised device. Five random seeds are used throughout.

\subsection{Variability-aware training} \label{sec:supp:variability:aware}

\cut{When the optimiser sees the actual device parameters during training, it compensates. Supplementary Table~\ref{tab:supp_variability_phase2} reports yacht [6,1,1] $R^2$ across 5~seeds per condition. $I_{\text{tr}}$ and $G_{\text{max}}$ perturbations produce near-nominal $R^2$. Three conditions show degradation: $\alpha_H$ ($R^2 = 0.778 \pm 0.176$), combined variability at $f = 10$\% ($0.839 \pm 0.275$), and $P_{\text{sat}}$ ($0.875 \pm 0.199$). The elevated standard deviations are driven in part by a single seed (seed~46) that converges to a narrow minimum sensitive to the sampled device parameters; excluding it brings the $\alpha_H$ mean to $0.857 \pm 0.086$ and the combined $f = 10$\% mean to $0.976 \pm 0.015$. Variability-aware training compensates device mismatch across all conditions, though $\alpha_H$ remains the axis where compensation is most difficult.}

\add{When the optimiser sees the actual device parameters during training, it compensates. Supplementary Table~\ref{tab:supp_variability_phase2} reports yacht [6,1,1] $R^2$ across 5~seeds per condition, using the sigmoid reparameterisation described in Methods. All eight perturbation conditions produce $R^2 \geq 0.955$, with the worst case being the combined ($f = 5$\%) condition ($R^2 = 0.955 \pm 0.062$). The linewidth enhancement factor $\alpha_H$ remains the most sensitive single-parameter axis ($R^2 = 0.961 \pm 0.047$), but variability-aware training compensates its effect to within 0.4~percentage points of the nominal $R^2$. Without the sigmoid reparameterisation (i.e.\ using softplus with post-hoc clamping), three conditions show catastrophic degradation ($\alpha_H$: $R^2 = 0.778 \pm 0.176$; combined $f = 10$\%: $0.839 \pm 0.275$; $P_{\text{sat}}$: $0.875 \pm 0.199$), driven by seeds that converge to narrow minima sensitive to the sampled device parameters. The sigmoid reparameterisation eliminates this failure mode by constraining each trainable parameter to its physical operating range through a smooth bijection (Methods), preventing the optimiser from reaching parameter-scale configurations that are fragile under device variation.}

\begin{table}[H]
\caption{Yacht [6,1,1] $R^2$ when training with device variability (5~seeds per condition\add{, sigmoid reparameterisation}). Device parameters are sampled per module at model initialisation; the optimiser compensates during training.}\label{tab:supp_variability_phase2}
\centering
\begin{tabular}{@{}lc@{}}
\toprule
Condition & Yacht [6,1,1] $R^2$ \\
\midrule
Nominal & $\cut{0.971 \pm 0.011}\add{0.965 \pm 0.041}$ \\
$G_{\text{max}}$ ablation & $\cut{0.981 \pm 0.009}\add{0.976 \pm 0.019}$ \\
$P_{\text{sat}}$ ablation & $\cut{0.875 \pm 0.199}\add{0.974 \pm 0.024}$ \\
$I_{\text{tr}} \pm 5$\% & $\cut{0.971 \pm 0.010}\add{0.986 \pm 0.003}$ \\
$I_{\text{tr}} \pm 10$\% & $\cut{0.973 \pm 0.010}\add{0.972 \pm 0.028}$ \\
$\alpha_H$ ablation & $\cut{0.778 \pm 0.176}\add{0.961 \pm 0.047}$ \\
Combined ($f = 5$\%) & $\cut{0.864 \pm 0.150}\add{0.955 \pm 0.062}$ \\
Combined ($f = 10$\%) & $\cut{0.839 \pm 0.275}\add{0.974 \pm 0.020}$ \\
\botrule
\end{tabular}
\end{table}

\subsection{Deployment without recalibration} \label{sec:supp:variability:deploy}

\begin{figure}[H] \centering
\includegraphics[width=\textwidth]{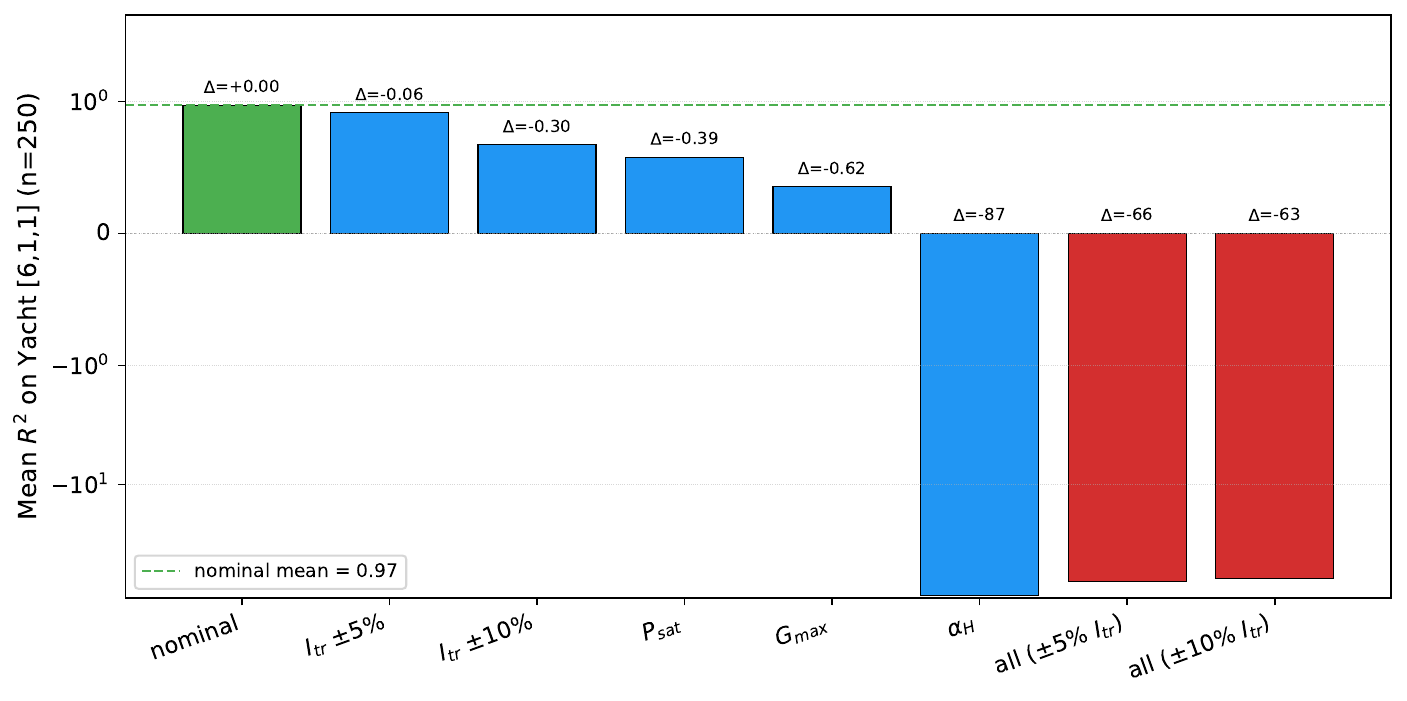}
\caption{Sensitivity of yacht [6,1,1] $R^2$ to single-parameter and combined device variation when deploying a nominally trained model without recalibration (250 evaluations per condition). $\alpha_H$, sampled over $\pm 40$\% of its nominal value, dominates all other parameter axes by two orders of magnitude. Sampling ranges for the remaining parameters are listed in Supplementary Table~\ref{tab:supp_variability_params}.}
\label{fig:supp_variability_ranking}
\end{figure}

\cut{The more demanding scenario is deploying a nominally trained model on hardware whose device constants differ from the training values. Supplementary Fig.~\ref{fig:supp_variability_ranking} ranks the eight conditions by mean $R^2$ across 250 evaluations (5~seeds $\times$ 50 Monte Carlo device draws). The $\alpha_H$ ablation is catastrophic: mean $R^2 = -86$, meaning predictions are worse than a constant-mean regressor on nearly all device draws. The combined conditions are similarly destroyed (mean $R^2 = -62$ to $-65$), dominated by the $\alpha_H$ contribution.}
\add{The more demanding scenario is deploying a nominally trained model on hardware whose device constants differ from the training values. Supplementary Fig.~\ref{fig:supp_variability_ranking} ranks the eight conditions by mean $R^2$ across 250 evaluations (5~seeds $\times$ 50 Monte Carlo device draws). The $\alpha_H$ ablation is catastrophic: mean $R^2 = -145$, meaning predictions are worse than a constant-mean regressor on nearly all device draws. The combined conditions are similarly destroyed (mean $R^2 = -98$ to $-101$), dominated by the $\alpha_H$ contribution. The sigmoid reparameterisation improves variability-aware training (Section~\ref{sec:supp:variability:aware}) but does not fix blind deployment, because the failure mechanism is a phase mismatch between the trained $\phi$ and the actual $\alpha_H$-dependent SOA phase shift.}

\begin{figure}[H] \centering
\includegraphics[width=\textwidth]{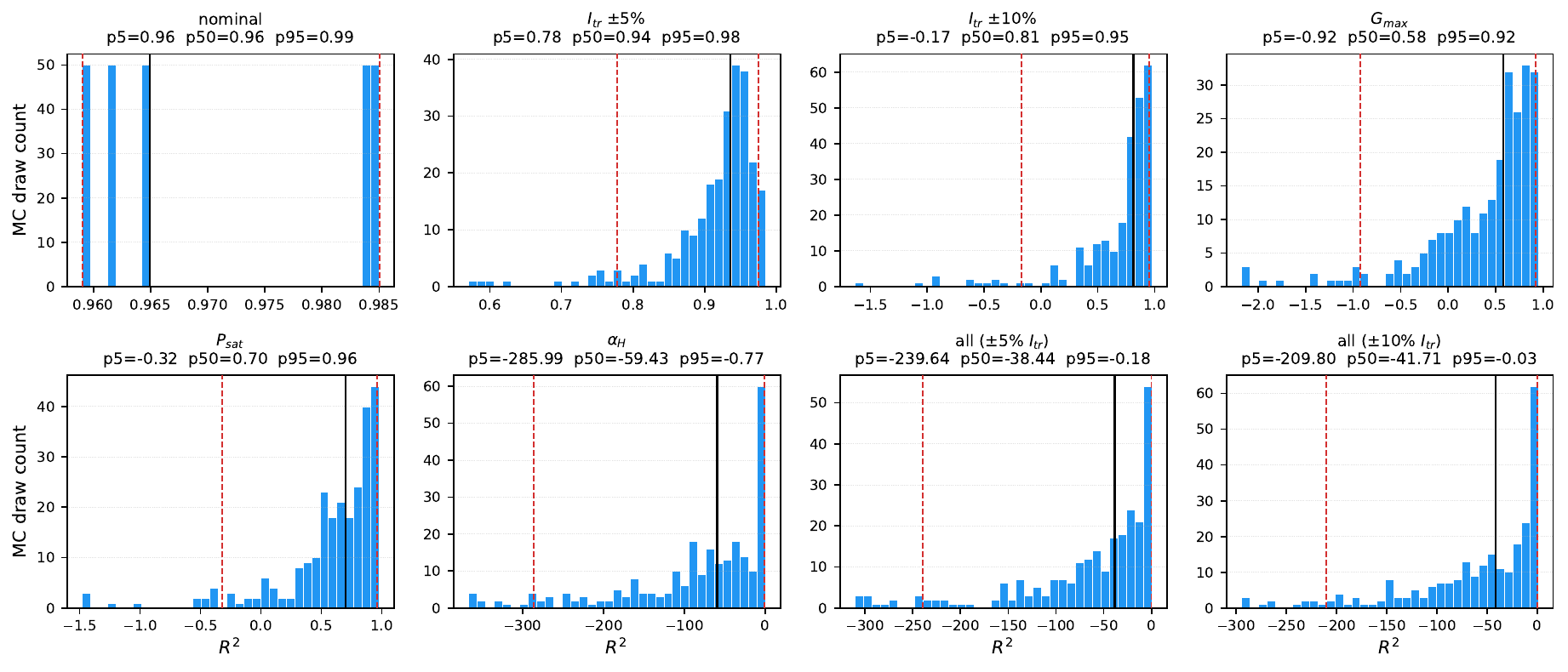}
\caption{Distribution of yacht [6,1,1] $R^2$ across 250 Monte Carlo device draws per condition, for a nominally trained model deployed without recalibration. Solid vertical lines mark the median; dashed red lines mark the 5th and 95th percentiles. The $\alpha_H$, combined ($\pm 5$\% $I_{\text{tr}}$), and combined ($\pm 10$\% $I_{\text{tr}}$) conditions produce uniformly negative $R^2$.}
\label{fig:supp_variability_histograms}
\end{figure}

\cut{Supplementary Fig.~\ref{fig:supp_variability_histograms} shows the full $R^2$ distributions. For $\alpha_H$ and the combined conditions, the 95th percentile remains negative, so the failure is not confined to a tail of extreme fabrication outcomes. $P_{\text{sat}}$, $G_{\text{max}}$, and $I_{\text{tr}} \pm 10$\% produce moderate degradation with heavy-tailed distributions: their median $R^2$ values (0.70, 0.58, 0.81) substantially exceed the means (0.58, 0.35, 0.67). $I_{\text{tr}} \pm 5$\% is the only perturbed condition that remains tolerable (mean $R^2 = 0.91$, $p_5 = 0.78$).}
\add{Supplementary Fig.~\ref{fig:supp_variability_histograms} shows the full $R^2$ distributions. For $\alpha_H$ and the combined conditions, the 95th percentile remains negative, so the failure is not confined to a tail of extreme fabrication outcomes. $P_{\text{sat}}$ and $G_{\text{max}}$ produce moderate degradation with heavy-tailed distributions: their median $R^2$ values (0.42, 0.12) substantially exceed the means ($-0.30$, $-0.63$). $I_{\text{tr}} \pm 5$\% is the only perturbed condition that remains tolerable (mean $R^2 = 0.89$, median $R^2 = 0.95$).}

\subsection{Physical origin and mitigation} \label{sec:supp:variability:discussion}

The $\alpha_H$ sensitivity has a direct physical origin. The linewidth enhancement factor enters the MZI interference condition through the gain-induced phase shift $\Delta\phi_{\text{SOA}} = -\alpha_H h/2$ (Methods). Training selects MZI phases $\phi$ jointly with the nominal $\alpha_H = 5$. Sampling $\alpha_H$ in $[3,\,7]$ shifts every interference condition by $\Delta\phi = -(\Delta\alpha_H / 2) \cdot h$. For a typical integrated gain $h \approx 3$, this produces phase errors of order $\pm 3$~rad, enough to move the operating point from a constructive to a destructive MZI fringe. In contrast, $P_{\text{sat}}$ and $G_{\text{max}}$ shift the gain saturation curve horizontally or vertically without altering the phase condition, preserving the qualitative shape of the transfer function.

The $\pm 40$\% $\alpha_H$ range used here ($[3,\,7]$ around nominal 5) is deliberately conservative. Telecom-grade SOAs are manufactured to tight specifications, and device-to-device $\alpha_H$ variation within a single product line is typically below $\pm 10$\%. The $[3,\,7]$ range spans the full spread reported across different SOA material systems and wavelength bands in the literature, not the variation expected within a batch of identical devices. The catastrophic failure observed under this range therefore represents a worst case that is unlikely to occur in practice with a single SOA product.

\cut{The mitigation is to characterise $\alpha_H$ per device. Standard gain--phase measurements extract $\alpha_H$ from the ratio of refractive index change to gain change, and are part of routine SOA characterisation (see also Supplementary Section~\ref{sec:supp:experimental}). Substituting the measured $\alpha_H$ into the forward model before training ensures that the optimised MZI phases account for the actual gain--phase coupling of each device. When per-device characterisation is available, variability-aware training produces robust solutions across all parameter axes at moderate cost on yacht (Supplementary Table~\ref{tab:supp_variability_phase2}).}
\add{Two complementary mitigations eliminate this failure mode. First, per-device characterisation of $\alpha_H$ from a standard gain--phase measurement (see Supplementary Section~\ref{sec:supp:experimental}) ensures that the forward model accounts for the actual gain--phase coupling. Second, the sigmoid reparameterisation (Methods) constrains each trainable parameter to its physical range through a smooth bijection, preventing the optimiser from reaching fragile parameter-scale configurations. With both mitigations in place, variability-aware training produces $R^2 \geq 0.955$ across all eight perturbation conditions (Supplementary Table~\ref{tab:supp_variability_phase2}), including the $\alpha_H$ ablation that was previously catastrophic.}

The variability study also bounds the effect of operating at different wavelengths. Shifting the source away from the SOA gain peak changes all four device constants simultaneously: the gain $G_{\text{max}}(\lambda)$ and noise figure NF($\lambda$) follow the material gain spectrum, $P_{\text{sat}}(\lambda)$ tracks the wavelength-dependent confinement factor, and $\alpha_H(\lambda)$ varies with the detuning from the band edge. For the BOA1554P, the datasheet specifies ${\sim}$3~dB gain variation across the 80~nm C-band (1520--1600~nm). This is comparable to the $G_{\text{max}}$ ablation range used here ($[32,\,35]$~dB, i.e.\ 3~dB one-sided), and falls within the regime where the nominally trained model remains functional (Supplementary Table~\ref{tab:supp_variability_phase2}). Multi-wavelength operation therefore requires per-channel characterisation and retraining with the wavelength-resolved Agrawal--Olsson parameters, but does not introduce a qualitatively new failure mode beyond those already captured by the single-parameter sweeps.

This study samples each module's device constants independently. Spatially correlated fabrication variation (e.g.\ a systematic carrier-density gradient across a photonic die) and temperature-induced parameter drift are not modelled. Independent sampling overestimates the spread that a real production batch would exhibit, so the deployment results represent a conservative bound on degradation.

\end{appendices}

\end{document}